\newif\ifjapanese

\ifjapanese
  \documentclass[a4j,11pt]{jreport}                  
  \renewcommand{\bibname}{参考文献}
  \newcommand{\acknowledgmentname}{謝辞}
\else
  \documentclass[a4paper,11pt]{report}
  
  \renewcommand{\bibname}{References}
  \newcommand{\acknowledgmentname}{Acknowledgment}
\fi

\usepackage{dcolumn}
\usepackage{makecell}
\usepackage{array}
\usepackage{booktabs}
\usepackage{thesis}
\usepackage{ascmac}
\usepackage[dvipdfmx]{graphicx}
\usepackage{multirow}
\usepackage{url}
\usepackage{here}
\usepackage{CJKutf8}
\usepackage{hyperref}
\usepackage{algorithm}
\usepackage{algpseudocode}
\usepackage{enumitem} 
\usepackage{amsmath}
\usepackage{listings}
\lstdefinelanguage{makefile}{
morekeywords={if,else,endif,define,endef,include},
sensitive=false,
morecomment=[l]{\#},
morestring=[b]"
}

\usepackage{lmodern} 
\usepackage[T1]{fontenc}
\usepackage{breakurl}
\usepackage{float} 
\usepackage{placeins}
\usepackage{subcaption}
\numberwithin{algorithm}{chapter}

\usepackage{indentfirst} 

\usepackage{amssymb} 

\bindermode  
\jclass  {修士論文}                             

\eclass  {Master's Thesis}                            
\etitle    {\shortstack[c]{Multi-target Coverage-based Greybox Fuzzing}}
\euniv{Institute of Information Security} 
\efaculty{Graduate School of Information Security} 
\eauthor  {5543501  Masami Ichikawa}                           
\eyear  {2025}                                        
\ekeyword  {Fuzzing, Coverage-based Greybox Fuzzing}          
\eproject{Supervisor Kuniyasu Suzaki}                 
\edate{January 2025}

\usepackage{listings}
\begin{document}

\ifjapanese
  \jmaketitle    
\else
  \emaketitle    
\fi

\begin{eabstract} 
In recent years, fuzzing has been widely applied not only to application software but also to system software, including the Linux kernel and firmware, and has become a powerful technique for vulnerability discovery. Among these approaches, Coverage-based greybox fuzzing, which utilizes runtime code coverage information, has become the dominant methodology.
Conventional fuzzing techniques primarily target a single software component and have paid little attention to cooperative execution with other software. However, modern system software architectures commonly consist of firmware and an operating system that operate cooperatively through well-defined interfaces, such as OpenSBI in the RISC-V architecture and OP-TEE in the ARM architecture. 

In this study, we investigate fuzzing techniques for architectures in which an operating system and firmware operate cooperatively. In particular, we propose a fuzzing method that enables deeper exploration of the system by leveraging the code coverage of each cooperating software component as feedback, compared to conventional Single-target fuzzing. 
To observe the execution of the operating system and firmware in a unified manner, our method adopts QEMU as a virtualization environment and executes fuzzing by booting the system within a virtual machine. This enables the measurement of code coverage across software boundaries. Furthermore, we implemented the proposed method as a Multi-target Coverage-based Greybox Fuzzer called \textit{MTCFuzz} and evaluated its effectiveness.

\end{eabstract}

\tableofcontents  
\listoffigures    
\listoftables    
\listofalgorithms 

\cleardoublepage
\begingroup
  \renewcommand{\thepage}{\roman{page}}
  \lstlistoflistings
\endgroup

\cleardoublepage

\pagenumbering{arabic}

\chapter{Introduction}
\section{Background and Motivation}
In recent years, fuzzing has been widely used not only in application software but also in lower-level areas such as the Linux kernel and firmware. In particular, Coverage-based Greybox Fuzzing(CGF), 
which uses code coverage at runtime as feedback, has become mainstream due to its efficient search capabilities. In open source software development, the OSS-FUZZ project~\cite{githubOssfuzzprojectsMaster} 
is continuously fuzzing over 1,000 projects as of March 2025. In the Linux kernel, fuzzing is performed continuously using syzkaller~\cite{githubGitHubGooglesyzkaller}. 
As with software testing, fuzzing can be classified into three types: whitebox-, blackbox-, and greybox fuzzing. In recent years, the greybox fuzzing method, which utilizes code coverage information during program execution, 
has become widely used. The main greybox fuzzers are AFL++~\cite{fioraldi2020aflplusplus}, which primarily targets application software, and libFuzzer~\cite{llvmLibFuzzerx2013}, which can fuzz a variety of 
applications and libraries. The syzkaller is a widely used fuzzer targeting operating systems (OS), and supports fuzzing not only for the Linux kernel but also for BSD-based OSes such as FreeBSD and NetBSD, 
as well as Android and Windows.

\begin{figure}[th]
\centering
\includegraphics[width=\textwidth]{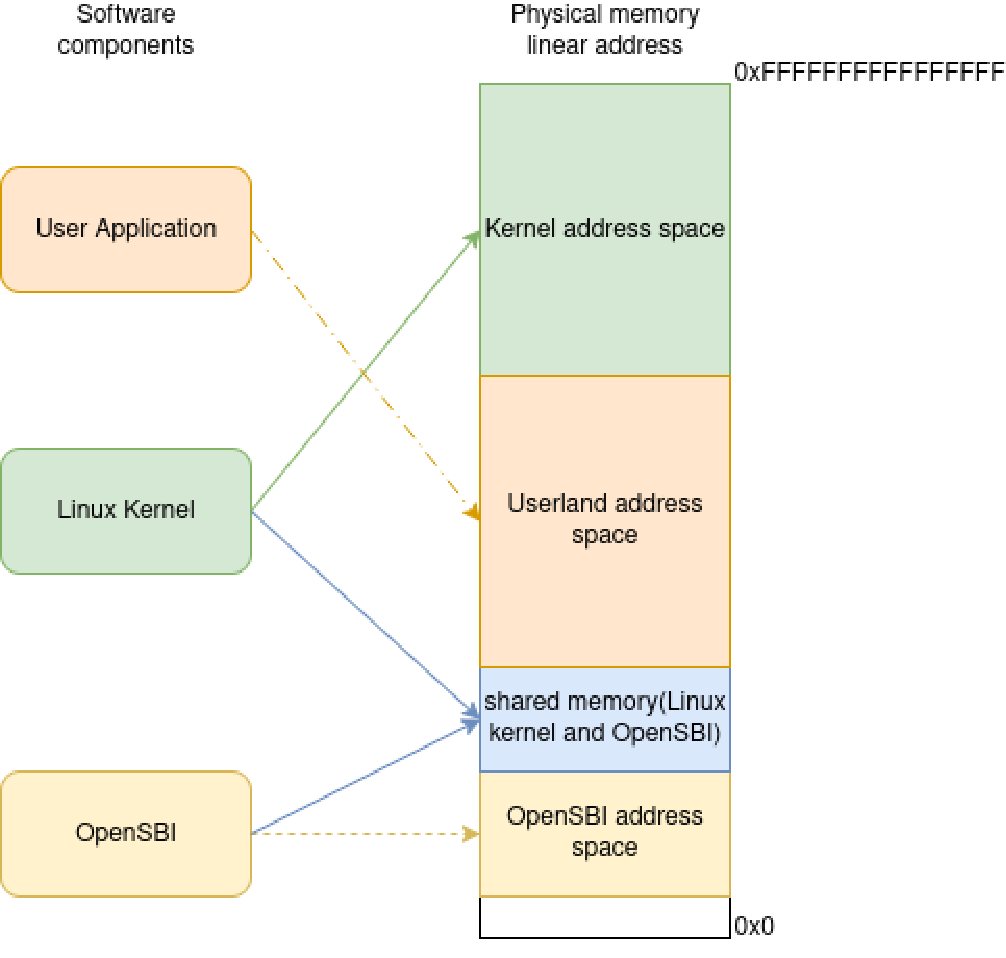}
\caption[
Memory layout on RISC-V
]{
Memory layout on RISC-V, where the User application, Linux kernel, and OpenSBI operate in isolated address ranges while interacting through a shared memory region.
}
\label{fig:opensbi-kernel-memory-layout.drawio.pdf}
\end{figure}

\begin{figure}[th]
\centering
\includegraphics[width=\textwidth]{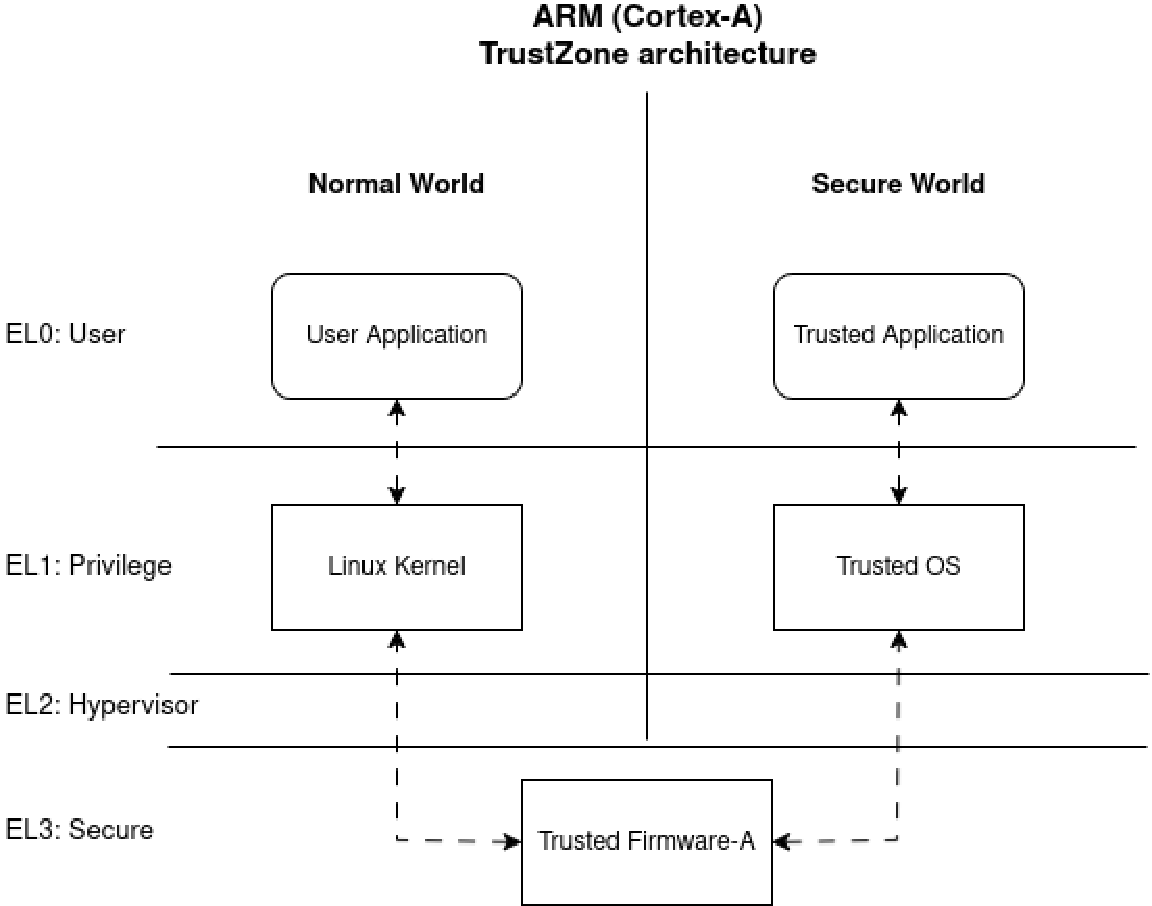}
\caption[
TrustZone for Cortex-A
]{
TrustZone architecture for Cortex-A, illustrating the separation between the Normal World and the Secure World.
}
\label{fig:secureworld.drawio.pdf}
\end{figure}

Fuzzers include not only tools developed for testing specific software, but also those that are capable of targeting a wide range of programs. However, in typical fuzzing tools, only one piece of software is selected as the fuzzing target. 
As a result, code coverage information from other software components that operate in cooperation with the target, such as database systems or firmware, is generally not taken into account. 
However, modern system software is generally configured so that multiple pieces of software operate cooperatively through specific interfaces. 
Examples include collaboration between Linux and OpenSBI in the RISC-V architecture~\cite{riscv_sbi_2_0}, the Normal World where Linux runs, and the Secure World where Trusted OS runs 
in TrustZone~\cite{armDocumentationx2013} in the ARM architecture. In this paper, TrustZone refers to TrustZone for Cortex-A. These architectures use the same physical memory as the OS, 
but memory is isolated by address ranges. For example, \autoref{fig:opensbi-kernel-memory-layout.drawio.pdf} shows the RISC-V memory layout, and the blue box shows the shared memory area between OpenSBI and the Linux kernel. 
\autoref{fig:secureworld.drawio.pdf} shows the TrustZone architecture. Each software component is isolated by its address range that is the same as RISC-V.
\autoref{tbl:optee-address-range} shows a classification of software components in the TrustZone environment based on address ranges extracted from fuzzing execution traces. 
Thus, it can be confirmed that multiple software components each occupy their own memory address ranges within the system memory.

\begin{table*}[th]
  \centering
  \caption[
    Software Classification Based on Address Ranges Observed in trace logs
  ]{
    Software Classification Based on Address Ranges Observed in ARMv8 TrustZone fuzzing trace logs.
  }
  \begin{tabular}{c|c}
    \hline
    \textbf{Address Ranges} & \textbf{Software} \\
    \hline
    0xe090000 -- 0xe0a00c4 & Trusted Firmware-A \\
    0xe100000 -- 0xe19f4c8 & Trusted OS \\
    0x40000400 -- 0x402ab510 & Trusted Application \\
    0xaaaaab0e55d4 -- 0xaaaae95ff1ac & Linux: User Application \\
    0xffff814d3ac4 -- 0xffff80008110f928 & Linux: Kernel Land \\
    \hline
  \end{tabular}
  \label{tbl:optee-address-range}
\end{table*}

Because the OS and firmware are developed independently and share the same physical memory through well-defined interfaces, incorrect input validation or inconsistent state handling at these interfaces can propagate across components. 
Since they are often developed by different teams or communities, gaps in the interpretation of specifications or assumptions about interface behavior can also become sources of subtle bugs. 
For example, data that should have been rejected by the firmware may instead be processed normally due to an implementation flaw, leading to subtle inconsistencies between OS and firmware states. 
Such issues have been observed in practice; for example, OpenSBI has fixed bugs related to feature value checking~\cite{opensbi_fwft_check_feature_value}. 
If such a bug directly causes a firmware crash, it can be detected by fuzzing the firmware. However, if the crash occurs later due to state inconsistency on the OS side, 
it may be difficult to detect using fuzzing that targets only the firmware.

In summary, conventional CGF mainly targets a single software component. Hence, it does not measure code coverage across software boundaries even in software where multiple components operate cooperatively. 
As a result, newly reached code coverage on the cooperating software side cannot be detected or utilized as feedback, and in modern system software with complex cooperative execution environments, 
this limitation leads to reduced fuzzing efficiency.

\FloatBarrier

\section{Objectives and Contributions}

In this study, our aim is to propose a fuzzing approach that simultaneously targets multiple cooperating software components and leverages their respective code coverage information as feedback, thereby enabling deeper system exploration than conventional methods.

\begin{itemize}
\item We enabled code coverage measurement without modifying the source code or binaries of the fuzzing target software.
\item We realized coverage tracing of executed basic blocks by QEMU.
\item We demonstrated the generality of our approach through empirical evaluation across multiple architectures (ARM and RISC-V).
\item We showed that using the code coverage of cooperating software components as feedback can improve the exploration capability of fuzzing.
\item We developed a prototype fuzzer called the \textit{Multi-target Coverage-based Greybox Fuzzer (MTCFuzz)} that implements the proposed approach~\cite{githubGitHubMasami256MTCFuzz}.
\item We discovered a previously unknown vulnerability, assigned to CVE-2025-40031~\cite{CVE-2025-40031}.
\item We received the OWS Research Award (2025) at the Computer Security Symposium 2025~\cite{iwsecOWS2025}.
\end{itemize}

\section{Organization of This Thesis}

The organization of this thesis is as follows.
Chapter \ref{ch:fuzzing} explains the fundamental technologies of fuzzing, including
fuzzing techniques, greybox fuzzing, and code coverage measurement,
and clarifies the limitations of conventional coverage-guided fuzzing.
Chapter \ref{ch:proposed-approch} describes the proposed \textit{Multi-target Coverage-based Greybox Fuzzing} approach, presents its design, and outlines the prototype
implementation, \textit{MTCFuzz}.
Chapter \ref{ch:Evalation} presents the experiments conducted to evaluate the
effectiveness of the proposed method and discusses the results.

Chapter \ref{ch:discussion} discusses the effectiveness and limitations of the proposed
approach based on the evaluation results.
Chapter \ref{ch:related-work} reviews related research on QEMU-based fuzzing frameworks
and fuzzers targeting kernels and Trusted OS.
Chapter \ref{ch:future-work} describes future work to further enhance and
 apply the proposed method.
Finally, Chapter \ref{ch:conclusion} concludes this thesis.

\chapter{Fuzzing}
\label{ch:fuzzing}
Fuzzing is a major software testing method these days. The technique of testing programs by providing random inputs was empirically known to be effective as early as the 1950s, when programs were entered using punch cards~\cite{secretsofconsultingFuzzTesting}. 
At that time, discarded punch cards were sometimes used as test inputs.

In 1990, Miller et al. conducted an empirical study to evaluate the robustness of UNIX utilities against random, invalid inputs, introducing the term ``fuzz``. 
In this paper, the term ``fuzz`` refers to random characters used for this testing method~\cite{miller1990empirical}. 
Motivated by accidental crashes caused by line noise during remote logins, they systematically fed random inputs to 88 UNIX utility programs across multiple systems. 
Their results showed that 25-33\% of the tested programs crashed or hung, revealing numerous latent bugs such as buffer overflows, unchecked return values, and pointer errors. 

\section{Fuzzing Techniques}

Fuzzing is a testing technique in which programs are exercised using randomly generated inputs. As such, it encompasses multiple approaches. Fuzzing can be categorized into several types. One common way to classify fuzzing is based on whether the internal structure of the software under test (SUT) is taken into account, similar to whitebox and blackbox testing in software testing. 
According to this classification, fuzzing can be divided into three types: whitebox-, blackbox-, and greybox fuzzing. 
In addition, in terms of test case generation methods, there are Mutation-based fuzzing, which mutates existing test case templates, and Generation-based fuzzing, which creates input from scratch. Several fuzzing approaches are named using 
Color-based terminology, such as black and white. 
These colors are used in the same sense as blackbox and whitebox testing in software testing. In this chapter, we describe the main techniques among them.

\vskip.5\baselineskip
\noindent\textbf{Whitebox fuzzing}:This approach analyzes the source code of the SUT and generates test cases using knowledge of internal structure and data structures in the SUT. Therefore, these internal knowledge can be used to guide test case generation, enabling higher code coverage. In whitebox fuzzing, fuzzers are implemented using symbolic execution~\cite{godefroid2008automated, stephens2016driller, yun2018qsym}. 

\vskip.5\baselineskip
\noindent\textbf{Blackbox fuzzing}:Unlike whitebox fuzzing,blackbox fuzzing does not analyze the SUT. Test cases are therefore generated on the basis of visible specifications, such as file formats, API interfaces, etc. The test result is judged by detecting crashes or abnormal signals. However, because blackbox fuzzing has no knowledge of the program’s internal structure, it faces challenges to achieve high code coverage~\cite{godefroid2007random}. For example, in the function shown in \autoref{code:if_condition}, the probability that $x$ equals $0xdeadbeef$, which is ${1}/{2^{32}}$. This is a very small probability, making it extremely unlikely that random mutations in blackbox fuzzing will exercise the path that evaluates the condition to true.

\begin{lstlisting}[
    breaklines=true,
    language=c,
    float=th, 
    caption={Simple if condition example},
    label={code:if_condition},
    numbers=left,
    basicstyle=\ttfamily\normalsize
]
bool f(u32 x)
{ 
  if (x == 0xdeadbeef)
    return true;
    
  return false;
}
\end{lstlisting}

\begin{algorithm}[th]
\centering
\caption{Basic Fuzzing Algorithm}
\label{algo:basic_fuzzing_code}
\begin{algorithmic}[1]
    \Require Seed input $S$
    \State $S_\dagger \gets \emptyset$
    \Repeat
        \State $s \gets \textsc{ChooseNext}(S)$
        \State $s' \gets \textsc{MutateInput}(s)$
        \State $t \gets \textsc{Execute}(s')$
        \If{$t$ is crashed}
            \State $S_\dagger \gets S_\dagger + s'$
        \EndIf
    \Until{Timeout reaches or all the seeds are checked}
    \Ensure Crash inputs $S_\dagger$
\end{algorithmic}
\end{algorithm}

\FloatBarrier

\vskip.5\baselineskip
\noindent\textbf{Mutation-based fuzzing}:In this approach, seeds that serve as templates for inputs are prepared, and during the mutation phase, these seeds are mutated to generate new test cases. Once the sample data are prepared, fuzzing can begin immediately. Although this approach allows data to be modified without requiring detailed knowledge of file formats or protocol specifications, the quality of generated test cases largely depends on the templates that were initially provided~\cite{miller2007analysis}.

\vskip.5\baselineskip
\noindent\textbf{Generation-based fuzzing}:Unlike Mutation-based fuzzing, Generation-based fuzzing does not rely on templates as the basis for test cases; instead, it constructs test cases from scratch. Therefore, it lacks the convenience of Mutation-based fuzzing, which can be easily executed once the sample data are prepared. However, because test cases are Generated-based on the specifications, Generation-based fuzzing can produce higher quality test cases than Mutation-based fuzzing. Miller et al. reported that, in their study on the mutation of the PNG file format, Mutation-based fuzzing achieved only 24\% of the code coverage obtained by Generation-based fuzzing on average~\cite{miller2007analysis}.

\vskip.5\baselineskip
\noindent\textbf{Snapshot-based fuzzing}:This approach saves the runtime state and restores that saved state after each test execution, thereby returning the execution environment to its pre-test condition. This enables efficient testing by avoiding the need to re-execute the setup steps required for each test. Moreover, some approaches do not use snapshots solely to save and restore the state, but instead mutate the saved state itself~\cite{shou2023ityfuzz}. Chaofan et al. implemented ItyFuzz~\cite{shou2023ityfuzz}, it freezes the program state and performs fuzzing using two complementary techniques: traditional mutations applied to transaction inputs and mutations applied directly to the saved state itself. Because the smart contracts targeted by ItyFuzz treats state as a critical component, mutating the state enables the discovery of state dependent bugs that cannot be revealed by input only mutations. 

\section{Greybox Fuzzing}
The Greybox fuzzing is major fuzzing technique nowadays. As its name suggests, this approach lies between whitebox and blackbox fuzzing. This approach does not analyze the SUT to the same extent as whitebox fuzzing; instead, it performs lightweight analysis or instrumentation and leverages that information to guide test case generation. Thus, while greybox fuzzing performs only lightweight analysis, the predominant technique is to measure code coverage and use it as feedback for test case generation. This methods that leverage code coverage in this manner are referred to as Coverage-based Greybox Fuzzing (CGF). 

The basic algorithm for blackbox fuzzing is given by the steps shown in \autoref{algo:basic_fuzzing_code}. The algorithm executes the tests as follows: 1) select a seed (line 3), 2) mutate the selected seed (line 4), 3) run the test (line 5), and 4) observe the execution and, if the program crashes, save the test case used (lines 6–7).

The corresponding algorithm for CGF is illustrated in \autoref{algo:greybox_fuzzing_code}. This algorithm is used by AFL~\cite{americanfuzzylop}. There are two differences from black box fuzzing: the first is the \textsc{AssignEnergy} function at line 4 and the second is the \textsc{IsInteresting} check at line 10.

\vskip.5\baselineskip
\noindent\textbf{AssignEnergy}: In blackbox fuzzing, no priorities are assigned to seeds, and all seeds are mutated uniformly. The \textsc{AssignEnergy} function assigns a priority, also known as energy, 
to each seed and determines how many mutations to perform for that seed according to its assigned energy (the assignment of priority at line 4 of \autoref{algo:greybox_fuzzing_code} and the loop at line 5 correspond to this behavior).
The \textsc{AssignEnergy} function assigns energy to each seed primarily based on the following criteria. It takes a seed as input and assigns energy according to that seed. In the AFL, 
the energy assigned by the \textsc{AssignEnergy} function is calculated based on execution time, basic block code coverage, and the time at which the seed was generated~\cite{bohme2016coverage}. 
In AFL, the functions \texttt{calculate\_score} and \texttt{fuzz\_one} correspond to \textsc{AssignEnergy}.

\begin{algorithm}[th]
\centering
\caption{Greybox Fuzzing Algorithm}
\label{algo:greybox_fuzzing_code}
\begin{algorithmic}[1]
    \Require Seed input $S$
    \State $S_\dagger \gets \emptyset$
    \Repeat
        \State $s \gets \textsc{ChooseNext}(S)$
        \State $p \gets \textsc{AssignEnergy}(s)$
        \For{$i$ from $1$ to $p$}
            \State $s' \gets \textsc{MutateInput}(s)$
            \State $t \gets \textsc{Execute}(s')$
            \If{$t$ is crashed}
                \State $S_\dagger \gets S_\dagger + s'$
            \ElsIf{\textsc{IsInteresting}($t$)}
                \State $S \gets S + s'$
            \EndIf
        \EndFor
    \Until{Timeout reaches or all the seeds are checked}
    \Ensure Crash inputs $S_\dagger$
\end{algorithmic}
\end{algorithm}

\FloatBarrier

\vskip.5\baselineskip
\noindent\textbf{IsInteresting}: The \textsc{IsInteresting} function determines whether a test case, although not causing a program crash, has produced an interesting test result. 
Here, “interesting” refers to cases such as reaching new code coverage or receiving a signal during execution. Test cases determined to be interesting by the \textsc{IsInteresting} function are added as new seeds. 
This enables, for example, to perform mutations on seeds that have reached new code coverage, which can in turn be expected to yield further coverage gains. 
Because blackbox fuzzing observes only crashes and does not monitor other behaviors, it cannot detect behavioral changes caused by different test cases. 
In contrast, CGF inspects the program’s runtime behavior through the \textsc{IsInteresting} function, enabling the generation of more effective test cases. 
In AFL, the functions \texttt{save\_if\_interesting} and \texttt{has\_new\_bits} in \texttt{afl-fuzz.c} correspond to \textsc{IsInteresting}.

\section{Code Coverage Measurement Techniques}
There are several methods to measure the coverage of the code. These include using compiler provided code coverage instrumentation~\cite{gnuGcovUsing, llvmSourcebasedCode}, 
extending compiler functionality to embed custom coverage instrumentation code~\cite{americanfuzzylop, fioraldi2020aflplusplus}, embedding coverage instrumentation code as part of the software itself~\cite{kernelKCOVCode}, 
measuring coverage through virtualized environments such as QEMU~\cite{githubAFLplusplusqemu_modeREADMEmdStable}, and using hardware tracking features to collect coverage information~\cite{schumilo2017kafl}. 
Each method offers its own advantages; however, when fuzzing multiple independently developed software components simultaneously—such as a Linux kernel and firmware the method for measuring code coverage 
must be unified between components. If the granularity of coverage differs among targets, it becomes difficult to integrate feedback between them.

\begin{lstlisting}[
    language=c, 
    float=th, 
    caption={AFL's code coverage instrumentation logic.},
    label={code:afl_instrumentation},
    numbers=left,
    breaklines=true,
    basicstyle=\ttfamily\normalsize
]
  cur_location = <COMPILE_TIME_RANDOM>;
  shared_mem[cur_location ^ prev_location]++; 
  prev_location = cur_location >> 1;
\end{lstlisting}

\begin{lstlisting}[
    language=c, 
    float=th, 
    caption={AFL's QEMU mode code coverage instrumentation logic.},
    label={code:afl_qemu_mode_instrumentation},
    numbers=left,
    breaklines=true,
    basicstyle=\ttfamily\normalsize
]
  if (block_address > elf_text_start && block_address < elf_text_end) {

    cur_location = (block_address >> 4) ^ (block_address << 8);
    shared_mem[cur_location ^ prev_location]++; 
    prev_location = cur_location >> 1;

  }
\end{lstlisting}

In this section, we describe three code coverage measurement techniques: SUT compile-time instrumentation and QEMU mode coverage instrumentation by AFL~\cite{americanfuzzylop}, and the KCOV feature in the Linux kernel~\cite{kernelKCOVCode}.

\FloatBarrier

\subsection{The code coverage measurement in the AFL}

\begin{figure}[tb]
\centering
\includegraphics[width=1.0\textwidth]{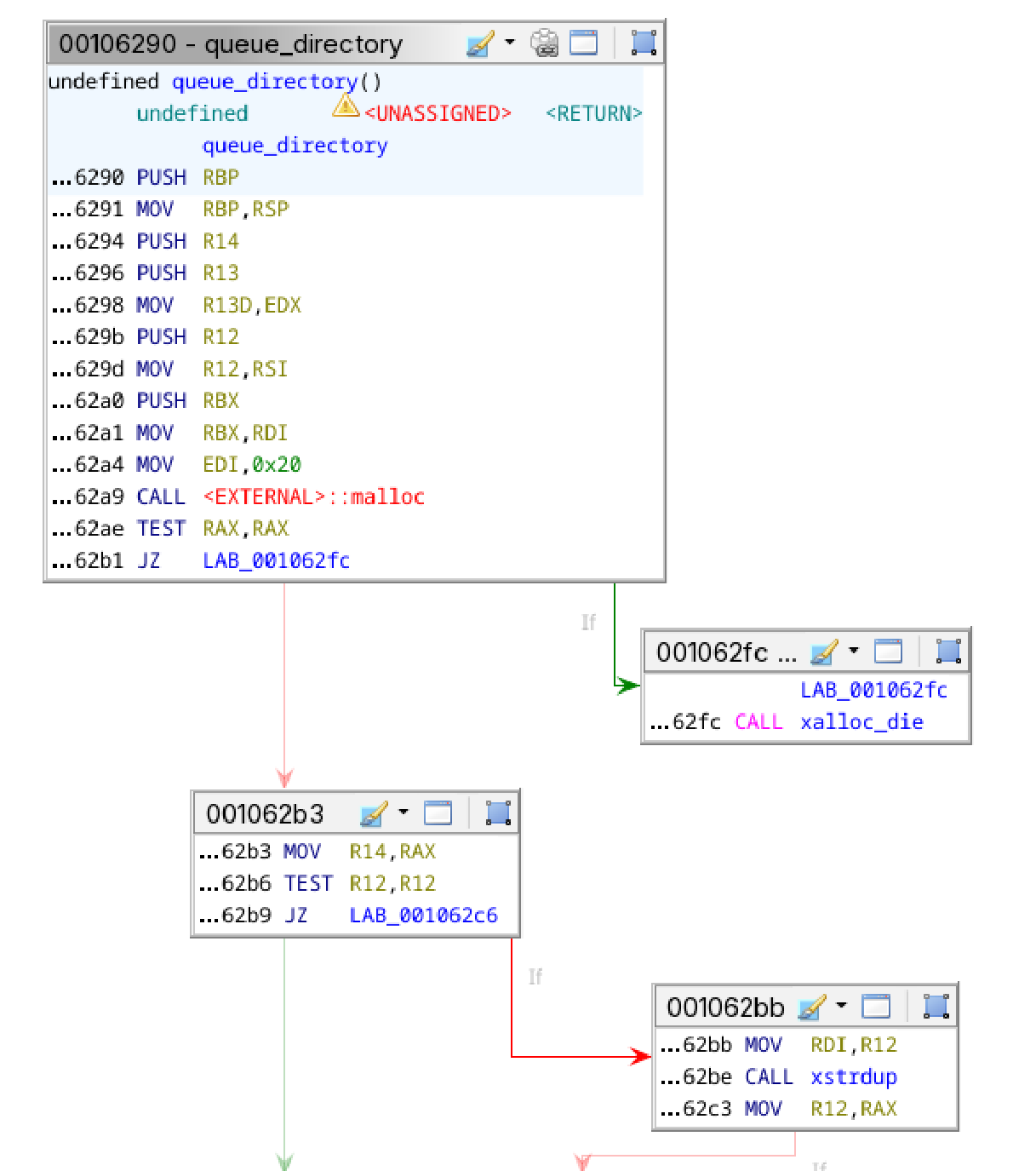}
\caption[
Example of basic blocks as displayed by Ghidra
]{
Example of basic blocks as displayed by Ghidra.
A basic block consists of a sequence of instructions ending at a branch instruction,
and control flow diverges to different successor blocks according to the branch.
}
\label{fig:ghidora-basic-block-sample.pdf}
\end{figure}

\begin{figure}[tb]
\centering
\includegraphics[width=\textwidth]{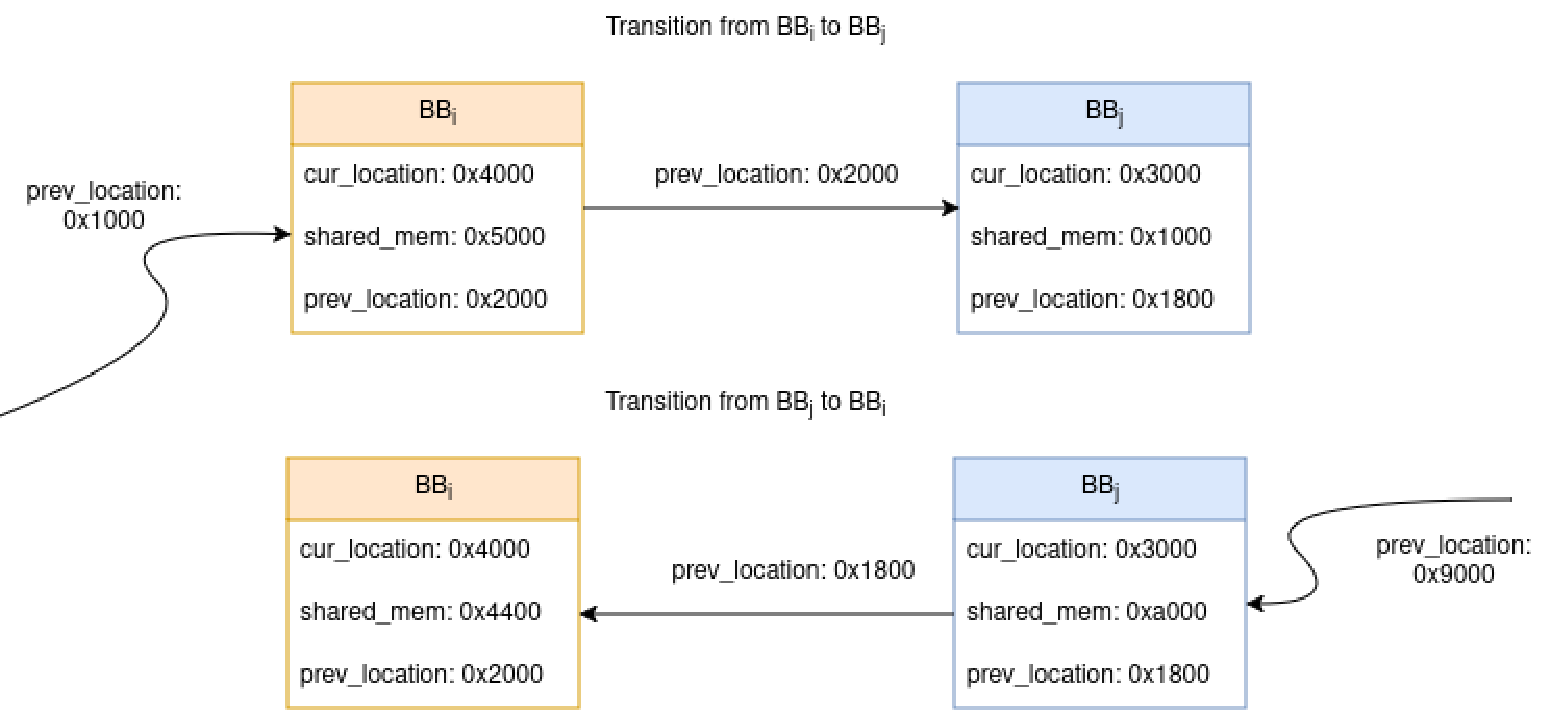}
\caption[
Example of basic block transitions in AFL
]{
Example of basic block transitions in AFL.
Transitions from $BB_i$ to $BB_j$ and from $BB_j$ to $BB_i$ are distinguished.
}
\label{fig:afl-bb-code-coverage.drawio.pdf}
\end{figure}

First, we explain how \texttt{AFL} embeds coverage instrumentation into the SUT. In \texttt{AFL}, a dedicated wrapper is used to compile the SUT, and during compilation, the code shown in \autoref{code:afl_instrumentation} is inserted in each basic block(BB). A basic block is a maximal sequence of instructions that is executed linearly, without any jumps or branches inside the block. An example of a basic block is shown in \autoref{fig:ghidora-basic-block-sample.pdf}. The value of \texttt{cur\_location} on line 1 is a unique identifier determined at the time of compilation for each basic block of the SUT. The variable \texttt{prev\_location} represents the basic block that was executed immediately prior to reaching the current one. In line 2, the XOR of \texttt{cur\_location} and \texttt{prev\_location} is used as an index in the shared memory array, and the number of transitions to the current BB is incremented at that index. In line 3, \texttt{prev\_location} is updated to \texttt{cur\_location} shifted right by one bit. This technique distinguishes the direction of transitions: moving from $BB_i$ to $BB_j$ and from $BB_j$ to $BB_i$ results in different indices, allowing the fuzzer to differentiate the transition directions.
For example, consider a transition from $BB_i$ to $BB_j$. Suppose that in $BB_i$, \texttt{cur\_location} is \texttt{0x4000} and \texttt{prev\_location} is \texttt{0x1000}. In this case, the index in the shared memory becomes \texttt{0x5000}, and \texttt{prev\_location} is updated to \texttt{0x2000}. Then, when execution reaches $BB_j$, where \texttt{cur\_location} is \texttt{0x3000}, the shared-memory index becomes \texttt{0x1000}. If the execution later changes from $BB_j$ to $BB_i$, \texttt{prev\_location} is updated to \texttt{0x1800} in $BB_j$. When the execution reaches $BB_i$, the shared-memory index becomes \texttt{0x4000\textasciicircum{}0x1800 = 0x4400}. Thus, the indices used for the transitions $BB_i \rightarrow BB_j$ and $BB_j \rightarrow BB_i$ differ, allowing the fuzzer to distinguish the direction of the transition. \autoref{fig:afl-bb-code-coverage.drawio.pdf} illustrates the transition states in this example.

Next, we describe how code coverage is measured in QEMU mode. Because QEMU mode does not require compiling the SUT as part of the fuzzing setup, no instrumentation code is embedded into the SUT itself.
Instead, coverage is measured within QEMU by treating a basic block as a unit of instrumentation~\cite{americanfuzzylop}. 
When QEMU performs emulation, it does not execute the target architecture’s machine code directly. 
Instead, it translates the code into an intermediate representation called Tiny Code Generator (TCG)~\cite{qemuDocumentationTCGQEMU}, 
using basic blocks as a translation unit. The method used in QEMU mode to measure code coverage is shown in \autoref{code:afl_qemu_mode_instrumentation}. 
In this approach, coverage is recorded only when the executed address lies within the text section of the ELF binary. This is because, in the ELF format, machine instructions are placed in the text section~\cite{xinuosSectionsx2014}. 
The subsequent operations on \texttt{cur\_location}, \texttt{shared\_mem}, and \texttt{prev\_location} follow the same procedure as described above. However, unlike compile time instrumentation, the value of \texttt{cur\_location} is derived by combining shifts and XOR operations on the basic block address to ensure its uniqueness.

\begin{table}[tb]
\centering
\caption{Common options for instrumentation}
\label{tab:trace-pc-cmp}
\begin{tabular}{lp{6cm}}
\hline
\textbf{Option} & \textbf{Inserted functions} \\
\hline
\texttt{-fsanitize-coverage=trace-pc}
 & \texttt{\_\_sanitizer\_cov\_trace\_pc} \\
\texttt{-fsanitize-coverage=trace-cmp}
 & \texttt{\_\_sanitizer\_cov\_trace\_cmp\{1,2,4,8\}} \newline
   \texttt{\_\_sanitizer\_cov\_trace\_const\_cmp\{1,2,4,8\}} \newline
   \texttt{\_\_sanitizer\_cov\_trace\_switch} \\
\hline
\end{tabular}
\end{table}

\begin{lstlisting}[
    language=makefile, 
    float=tb, 
    caption={Compiler flags for KCOV instrumentation in the Linux kernel},
    label={code:linux-makefile-kcov},
    numbers=left,
    breaklines=true,
    basicstyle=\ttfamily\normalsize
]
kcov-flags-y					                += -fsanitize-coverage=trace-pc
kcov-flags-$(CONFIG_KCOV_ENABLE_COMPARISONS)	+= -fsanitize-coverage=trace-cmp
\end{lstlisting}

\begin{lstlisting}[
    language=c, 
    float=tb, 
    caption={Example output of KCOV in trace-pc mode.},
    label={code:linux-kcov-output},
    numbers=left,
    basicstyle=\ttfamily\normalsize,
    breaklines=true
]
0xffffffff81d88e26
0xffffffff81d88c20
0xffffffff81e1940e
0xffffffff81e19497
0xffffffff81e195fe
0xffffffff81d88d6f
\end{lstlisting}

\FloatBarrier

\subsection{The code coverage measurement in the Linux kernel}
\texttt{KCOV} is a feature of the Linux kernel designed to support CGF. Unlike \texttt{AFL}, \texttt{KCOV} does not embed its own instrumentation code. Instead, 
it relies on the sanitizer coverage features provided by GCC~\cite{gnuInstrumentationOptions} and LLVM~\cite{llvmSanitizerCoveragex2014} to insert coverage hooks.
With these Compiler-based mechanisms, the compiler determines the instrumentation points and, upon reaching each point, invokes functions defined by the sanitizer specification. 
The function names and interfaces invoked at these points are standardized, and developers must implement their handlers according to this specification. Moreover, many of the sanitizer defined functions are shared between GCC and LLVM. 
The common interfaces available in both compilers are summarized in \autoref{tab:trace-pc-cmp}. 
When the \texttt{CONFIG\_KCOV} option is enabled in the Linux kernel configuration, the compiler is invoked with the sanitizer coverage options shown in \autoref{code:linux-makefile-kcov}. The compiler option \texttt{-fsanitize-coverage} accepts either \texttt{trace-pc} or \texttt{trace-cmp} as its argument. The \texttt{trace-pc} option traces only program counter addresses, whereas the \texttt{trace-cmp} option additionally traces values involved in conditional branches, 
including operands used in comparisons. In the Linux kernel, \texttt{trace-pc} is used as the default coverage instrumentation method, and enabling \texttt{CONFIG\_KCOV\_ENABLE\_COMPARISONS} allows the kernel to 
additionally trace comparison data observed during conditional branches. \texttt{KCOV} writes code coverage data to the file \texttt{/sys/kernel/debug/kcov}, and the fuzzer reads this file to obtain and analyze 
coverage information during fuzzing. The output format of \texttt{KCOV} is simple; in \texttt{trace-pc} mode, it consists of a sequence of addresses, as shown in \autoref{code:linux-kcov-output}.

\section{Background and Limitations of Coverage-Guided Fuzzing}

In this chapter, we discuss the major fuzzing techniques and provide an overview of CGF, which has become the mainstream approach in recent years. In particular, we explain the characteristic fuzzing algorithm of CGF and describe core code coverage measurement techniques, focusing on the \texttt{AFL} instrumentation and the \texttt{KCOV} feature in the Linux kernel.
Although each of these techniques has its own advantages, they are fundamentally designed for Single-target fuzzing. As a result, it is difficult to apply a uniform coverage measurement method across software components that are developed and built independently, such as an operating system and firmware. Furthermore, measuring coverage across software boundaries, for example, during transitions from OS to firmware is inherently challenging.
Therefore, in system configurations where independently developed components, such as the operating system and firmware, operate cooperatively, it is essential to employ a unified coverage measurement approach that does not depend on the build process of each component. Building on this understanding, the next chapter introduces our proposed method, which leverages unified coverage feedback across multiple cooperating software components.
Based on this understanding, the next chapter introduces our proposed \textit{Multi-target Coverage-based Greybox Fuzzing} method.

\chapter{Proposed Approach}
\label{ch:proposed-approch}
In the previous chapter, we described the general techniques used in CGF. As discussed earlier, CGF relies heavily on code coverage measurement; however, existing methods measure coverage for only a single software component, either by embedding instrumentation at compile time or by collecting coverage information during fuzzing execution. Consequently, these approaches do not consider how to measure code coverage in a unified manner across multiple software components that operate cooperatively within a system.

In this chapter, we describe the design and implementation of our proposed method, \textit{Multi-target Coverage-based Greybox Fuzzing}, along with its prototype fuzzer, \textit{MTCFuzz}, which we developed to realize this approach.

\section{Multi-target Coverage-based Greybox Fuzzing Method}

\begin{figure}[tb]
\centering
\includegraphics[width=\textwidth]{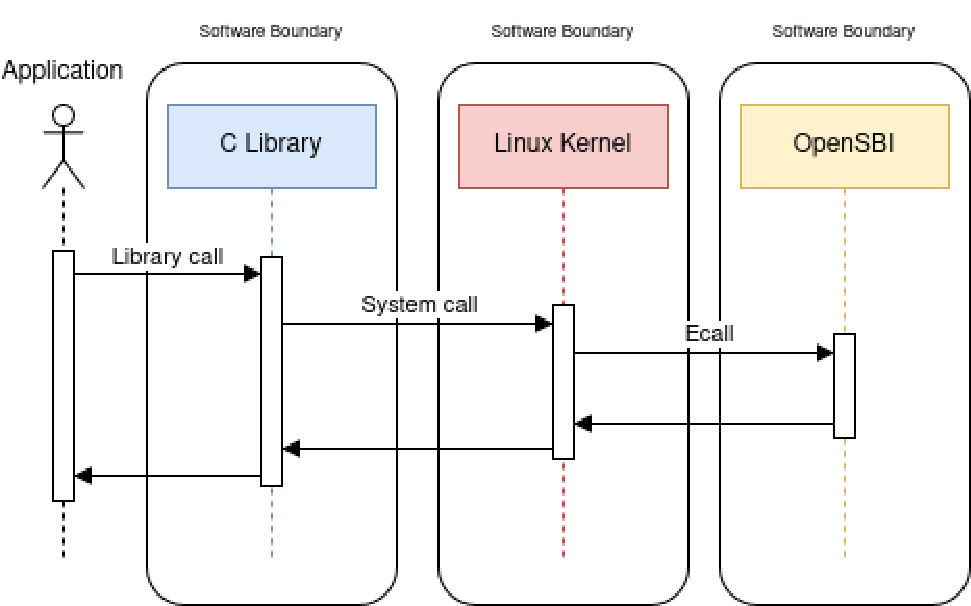}
\caption[
Execution flow across software boundaries
]{
Execution flow across software boundaries, where execution propagates from applications to the OS and firmware.
}
\label{fig:multi-software-boundaries}
\end{figure}

In our proposed method, \textit{Multi-target Coverage-based Greybox Fuzzing}, tracing the execution of multiple software components is essential. 
\autoref{fig:multi-software-boundaries} illustrates a scenario in which an application invokes a function in a library, the C library delegates processing to the Linux kernel via a system call, the Linux kernel further delegates processing to OpenSBI, and the result is returned to the application in reverse order. 
In the example, software boundaries exist not only at the application level, but also across the C library, the Linux kernel, and the OpenSBI.
Although these software components are developed independently and use different build processes, they all run within a single system environment.
Therefore, if the behavior of the software running on this system can be observed, 
it is possible to trace execution across software boundaries. Based on the observation, our proposed method adopts QEMU~\cite{qemuQEMU} as the execution
environment and traces the execution of machine-level instructions within QEMU, thereby allowing execution to be traced across multiple software boundaries. 
Unlike regular CGF, which measures coverage for only a single software component, our method unifies the coverage of multiple cooperating components and uses it as a single feedback signal for fuzzing.

\begin{figure}[tb]
\centering
\includegraphics[width=\hsize]{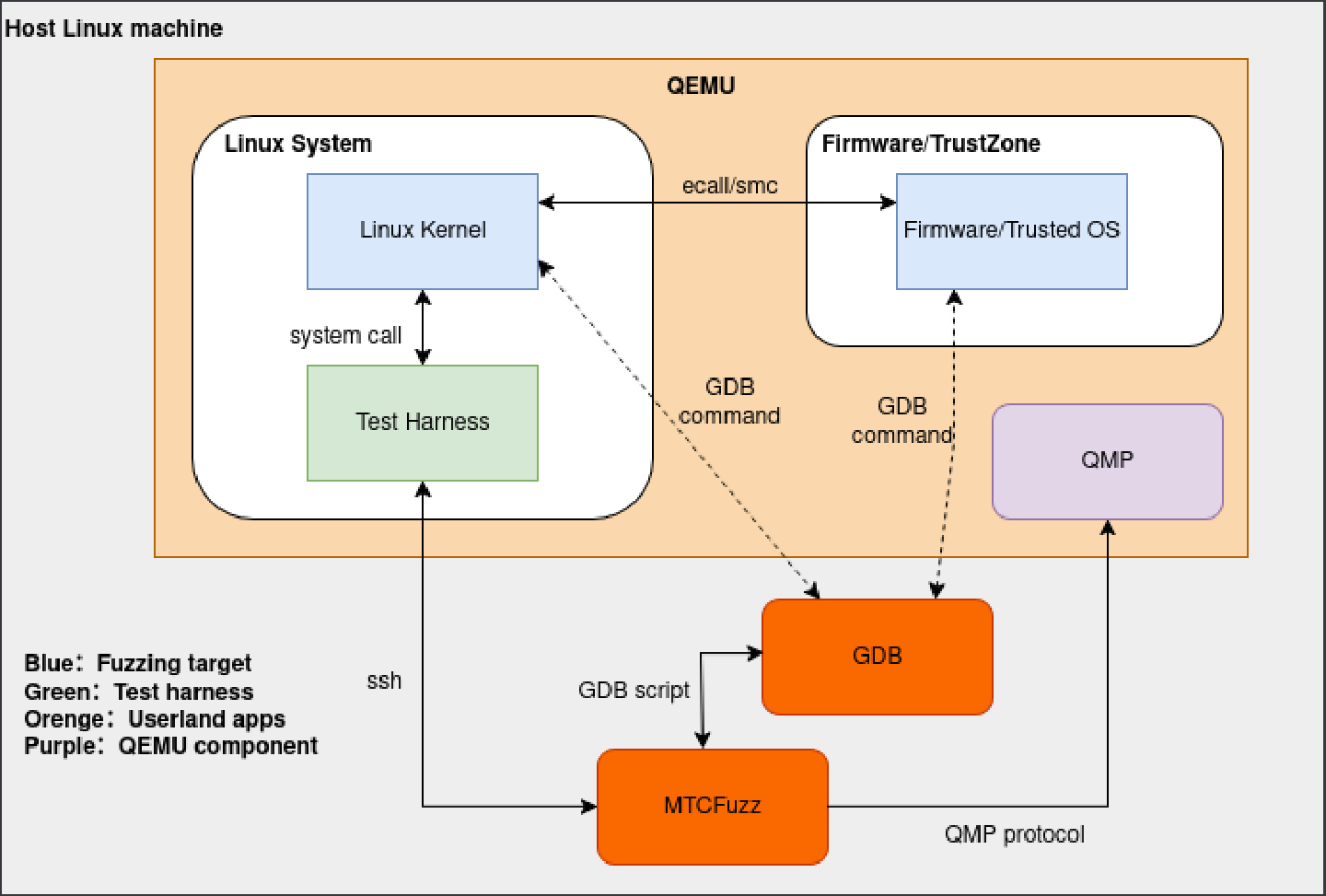}
\caption{Architecture of MTCFuzz based on QEMU.}
\label{fig:MTCFuzz-architecture}
\end{figure}

\vskip.5\baselineskip
\noindent\textbf{Unified code coverages:}

In the proposed method, a single input triggers execution across multiple software components. Consequently, the feedback used for test case evaluation must also be unified across all targets. 
Let $C(T_{\text{kernel}}^{i})$ and $C(T_{\text{fw}}^{j})$ denote the coverage observed in each execution unit. 
As shown in \autoref{eq:multi-coverage}, we define the unified multi-target coverage as
\begin{equation}
\label{eq:multi-coverage}
C_{\text{multi}} =
\left(\bigcup_{i=1}^{N_k} C(T_{\text{kernel}}^{i})\right)
\cup
\left(\bigcup_{j=1}^{N_o} C(T_{\text{fw}}^{j})\right).
\end{equation}

Here, the subscripts $\text{kernel}$ and $\text{fw}$ denote the Linux kernel and firmware (e.g., OpenSBI, Trusted OS, Trusted Application), respectively. 
The superscripts $i$ and $j$ represent the indices of the execution units observed during a single fuzzing execution, 
where $N_k$ and $N_o$ denote the total number of execution units observed in the kernel and firmware, respectively.

In summary, instead of evaluating each component independently, our proposed method treats the combined coverage space of the OS and firmware as a single exploration domain.

\begin{lstlisting}[
    float=tb, 
    caption={Address filters setting},
    label={code:address-filters},
    numbers=left,
    basicstyle=\ttfamily\normalsize,
    breaklines=true
]
    "address_filters": {
        "kernel": [
            {
                "name": "riscv_ext_d_validate",
                "lower": "0xffffffff80010764",
                "upper": "0xffffffff80010772"
            },
            ...
        ],
        "firmware": [
           ...
        ]
\end{lstlisting}

\vskip.5\baselineskip
\noindent\textbf{Measurement of code coverage:}
In each test execution, the code coverage observed during the execution of a test case is evaluated after the test case has finished. In \textit{MTCFuzz}, QEMU-based instrumentation is started at the beginning of each test 
execution and terminated when the test completes. The instrumentation results are recorded to a file. The sequence of recorded code coverage from the beginning to the end represents the execution flow of the code for that test case.
During fuzzing, the Linux kernel executes not only the target functionality under test but also unrelated components such as the scheduler and memory management. As a result, if the coverage of such an irrelevant code is 
included in the measurement, \textsc{IsInteresting} may return undesirable judgments. To address this issue, \textit{MTCFuzz} provides an address based filtering mechanism. The address filters are specified in a JSON file, 
as shown in \autoref{code:address-filters}. In the current implementation at the time of writing this paper, the filter configuration supports two types of target: \texttt{kernel} and \texttt{firmware}. The \texttt{kernel} 
filter is used to specify the address ranges of the Linux kernel, while the \texttt{firmware} filter is used to specify the address ranges of firmware components, such as OpenSBI, Trusted OS, and Trusted Applications. 
The code coverage data removed by filtering is concatenated into a single string and hashed, and the resulting hash value is recorded as the result of the test execution for the corresponding seed. By managing the execution 
flow as a whole rather than individual traced addresses, this approach enables differences in loop iteration counts, such as those in \textsc{for} and \textsc{while} loops, to be taken into account. The computed hash value 
is stored as part of the seed’s metadata, and the number of times the same code coverage has been observed in different seeds is aggregated. This count is then used by the power scheduler when calculating the energy assigned to each seed.

\section{Design of MTCFuzz}

In this section, we describe the implementation of \textit{MTCFuzz}. The fuzzing execution environment of our proposed method is shown in \autoref{fig:MTCFuzz-architecture}. When performing fuzzing with \textit{MTCFuzz}, 
the main software components include the fuzzer itself (\textit{MTCFuzz}), the target operating system (e.g., Linux), and the target firmware (e.g., OpenSBI or OP-TEE environments such as Trusted OS). \textit{MTCFuzz} is 
responsible for managing the entire fuzzing process. The management of the fuzzing process includes launching QEMU, saving and loading snapshots, seed mutation, code coverage checking, execution monitoring, and crash detection.

\textit{MTCFuzz} is designed to fuzz low-level software components such as kernels and firmware. If the fuzzing process were managed on Linux inside the virtual machine, a crash of the Linux kernel would cause the process 
responsible for managing the fuzzing campaign itself to stop, making it impossible to continue fuzzing. Therefore, in \textit{MTCFuzz}, the management of the fuzzing process is not performed inside the virtual machine, 
but is instead implemented as a process running on the host operating system. While the fuzzing process itself is managed by \textit{MTCFuzz}, the execution of the test harness must be performed inside the virtual machine. 
For this reason, \textit{MTCFuzz} transfers the test harness into the guest operating system during the setup phase using mechanisms such as \texttt{scp} or the \texttt{9p} file system, and then launches the test harness 
via \texttt{ssh}. Consequently, the ability to communicate with the guest operating system via \texttt{ssh} is a mandatory requirement for \textit{MTCFuzz}. 

\section{Implementation of MTCFuzz}

\vskip.5\baselineskip
\noindent\textbf{Measuring code coverage in QEMU:}

QEMU provides a debugging feature that traces the machine instructions executed~\cite{qemuEmulationx2014}; however, this feature continuously records the execution from the start to the end of QEMU. As a result, 
it is not suitable for selective instrumentation, only for the execution of individual test cases. When measuring code coverage in QEMU, it is therefore essential to be able to arbitrarily control the start and end of the measurement. 
In addition, QEMU must send the measurement results to a destination that can be read by the fuzzer. To meet these requirements, we extend the QEMU QMP API~\cite{qemuDocumentationQMPQEMU} to \textit{MTCFuzz} by adding new commands 
to starting and stopping coverage tracking. The definitions of these APIs are shown in \autoref{code:qmp-apis}. The API for starting code coverage measurement is \texttt{mtcfuzz-trace-start}, which takes a file name as 
the \texttt{filename} parameter and writes the trace log to the specified file. To stop the measurement, \texttt{mtcfuzz-trace-stop} is executed.

\textit{MTCFuzz} performs instrumentation at the timing when QEMU translates guest machine instructions into TCG. This translation is performed on a basic block basis~\cite{qemuDocumentationTCGQEMU}, 
and therefore the tracing granularity in \textit{MTCFuzz} corresponds to the level of QEMU basic blocks. Moreover, since TCG translation is used for all architectures supported by QEMU, the instrumentation mechanism can be implemented without 
introducing architecture-specific code, enabling a simple and portable design. The implementation of the QEMU instrumentation code is shown in patch format in \autoref{code:qemu-implementation-code}. The changes made to the 
existing source code consist only of adding calls to \texttt{mtcfuzz\_record\_tb\_exec} in two functions and therefore have minimal impact on the original codebase. As a result, even when upgrading the QEMU version, 
it is relatively easy to keep the implementation up to date. \texttt{mtcfuzz\_record\_tb\_exec} is the function shown in \autoref{code:qemu-implementation-mtcfuzz-record-tb-exec}.

\begin{lstlisting}[
    float=tb, 
    caption={Definition of APIs for code coverage instrumentation},
    label={code:qmp-apis},
    numbers=left,
    basicstyle=\ttfamily\normalsize,
    breaklines=true
]
##
# @mtcfuzz-trace-start:
#
# Start a fuzzing trace.
#
# @filename: Name of the file to store the trace.
#
# Since: 10.0
#
# .. qmp-example::
#
#     -> { "execute": "mtcfuzz-trace-start",
#          "arguments": { "filename": "trace.log" } }
#     <- { "return": {} }
##
{ 'command': 'mtcfuzz-trace-start',
  'data': { 'filename': 'str' }}

##
# @mtcfuzz-trace-stop:
#
# Stop a fuzzing trace.
#
# Since: 10.0
#
# .. qmp-example::
#
#     -> { "execute": "mtcfuzz-trace-stop" }
#     <- { "return": {} }
##
{ 'command': 'mtcfuzz-trace-stop'}
\end{lstlisting}

\begin{lstlisting}[
    float=tb, 
    caption={Implementation of instrumentation code in TCG translation process},
    label={code:qemu-implementation-code},
    numbers=left,
    basicstyle=\ttfamily\footnotesize,
    breaklines=true
]
diff --git a/accel/tcg/cpu-exec.c b/accel/tcg/cpu-exec.c
index 8163295f34..b20bd8f11d 100644
--- a/accel/tcg/cpu-exec.c
+++ b/accel/tcg/cpu-exec.c
@@ -41,6 +41,8 @@
 #include "tb-context.h"
 #include "internal-common.h"
 #include "internal-target.h"
+#include "trace/mtcfuzz_trace.h"
+
 
 /* -icount align implementation. */
 
@@ -603,6 +605,7 @@ void cpu_exec_step_atomic(CPUState *cpu)
 
         cpu_exec_enter(cpu);
         /* execute the generated code */
+        mtcfuzz_record_tb_exec((uint64_t) pc);
         trace_exec_tb(tb, pc);
         cpu_tb_exec(cpu, tb, &tb_exit);
         cpu_exec_exit(cpu);
@@ -904,6 +907,7 @@ static inline void cpu_loop_exec_tb(CPUState *cpu, TranslationBlock *tb,
                                     vaddr pc, TranslationBlock **last_tb,
                                     int *tb_exit)
 {
+    mtcfuzz_record_tb_exec((uint64_t) pc);
     trace_exec_tb(tb, pc);
     tb = cpu_tb_exec(cpu, tb, tb_exit);
     if (*tb_exit != TB_EXIT_REQUESTED) {
\end{lstlisting}

\begin{lstlisting}[
    float=tb, 
    caption={Implementation of writting trace log to file},
    label={code:qemu-implementation-mtcfuzz-record-tb-exec},
    numbers=left,
    basicstyle=\ttfamily\footnotesize,
    breaklines=true
]
void mtcfuzz_record_tb_exec(uint64_t pc)
{
    qemu_mutex_lock(&mtcfuzz_tb_trace_lock);
    
    if (!mtcfuzz_tb_trace_fp) {
        qemu_mutex_unlock(&mtcfuzz_tb_trace_lock);
        return;
    }

    fprintf(mtcfuzz_tb_trace_fp, "0x%" PRIx64"\n", pc);
    qemu_mutex_unlock(&mtcfuzz_tb_trace_lock);
}
\end{lstlisting}

\vskip.5\baselineskip
\noindent\textbf{The code coverage measurement and filtering:}

The code coverage traced by QEMU is filtered in \textit{MTCFuzz}.
Filters are described as a list in JSON format in a configuration file, as shown in \autoref{code:address-filters}.
The removal of unnecessary data from the traced coverage is performed by the \texttt{analyze\_coverage} procedure.
The algorithm of \texttt{analyze\_coverage} is shown in \autoref{algo:coverage_analysis}.
In this process, among the measured code coverage, if an address belongs to the target of measurement, the number of executions of the address is recorded in $K_{\text{cov}}$ (Linux kernel coverage) or $F_{\text{cov}}$ (firmware coverage).

During the filtering process, the target address is first checked to determine whether it is already registered in $K_{\text{cov}}$ or $F_{\text{cov}}$.
If the address has already been registered, its count is incremented.
If the address is not registered in either list, \texttt{AddrInFilters} is invoked to determine whether the address is included in the filtering list.
If this check is performed naively, the computational complexity required to examine a single address is $O(N)$, and if there are $M$ addresses to be examined, the total complexity becomes $O(MN)$.
Therefore, as the number of filtering targets increases, the performance of the filtering process is expected to degrade.
To address this issue, \textit{MTCFuzz} sorts the filter list in ascending order and applies binary search to the filtering procedure, thereby reducing the complexity of examining a single address to $O(\log N)$.
Consequently, the total complexity required to examine all traced addresses becomes $O(M \log N)$.
In addition, the measured addresses are stored in $K_{\text{cov}}$ or $F_{\text{cov}}$, and before executing the binary search, the algorithm checks whether the target address is already registered in one of these data structures (implemented as python dictionary type).
This membership test requires $O(1)$ expected time.
Therefore, in the filtering process, the best-case complexity is $O(1)$, while the worst-case complexity is $O(M \log N)$.
In the early stage of fuzzing campaign, new code coverage is likely to be discovered frequently.
Thus, a large number of new code coverage points are detected shortly after fuzzing begins. therefore, for a certain period after the start of fuzzing, the computational complexity remains close to $O(M \log N)$.
However, as the number of test executions increases, the number of newly discovered coverage points decreases, and the slope of the curve becomes gradually smaller.
Consequently, while the computational complexity of the filtering process is $O(M \log N)$ in the initial phase of fuzzing, it approaches $O(1)$ as the fuzzing.
Let $N$ be the number of filters, $K$ be the total number of distinct addresses that have appeared at least once so far, and $T$ be the total number of address checks performed throughout the fuzzing campaign (i.e., the sum over all campaigns, all tests, and all PCs).
Under these definitions, the total cost ($TotalCost$) can be expressed as shown in \autoref{eq:totalcost}, and the amortized complexity can be expressed as shown in \autoref{eq:average-cost}.
As fuzzing progresses, the discovery rate of new coverage decreases, and the increase in $T$ becomes dominant over that of $K$.
As a result, the ratio $K/T$ converges to zero and the computational complexity approaches $O(1)$.
This functionality is not merely a supporting mechanism for collecting code coverage, but a core component that directly influences fuzzing efficiency, and therefore plays an essential role in CGF.

\begin{equation}
\label{eq:totalcost}
\text{TotalCost} = O(T + K \log N),
\end{equation}

\begin{equation}
\label{eq:average-cost}
O\!\left(\frac{T + K \log N}{T}\right)
= O\!\left(1 + \frac{K}{T} \log N\right).
\end{equation}

\begin{algorithm}[th]
\centering
\caption{Coverage Classification for Kernel and Firmware Address Spaces}
\label{algo:coverage_analysis}
\begin{algorithmic}[1]
    \Require Covered PC list (hex strings) $C$
    \Require Kernel coverage map $K_{\text{cov}}$, firmware coverage map $F_{\text{cov}}$
    \Require Kernel filters $K_{\text{filter}}$, firmware filters $F_{\text{filter}}$
    \Require Kernel start list $K_{\text{starts}}$, firmware start list $F_{\text{starts}}$
    \Ensure Updated coverage maps $K_{\text{cov}}, F_{\text{cov}}$
    \State $A \gets \emptyset$ \Comment{List of all observed PCs}
    \State $K_{\text{found}} \gets \textbf{false}$
    \State $F_{\text{found}} \gets \textbf{false}$
    \ForAll{$pc_{hex} \in C$}
        \State $pc \gets \textsc{HexToInt}(pc_{hex})$
        \If{$pc \in K_{\text{cov}}$}
            \State $K_{\text{cov}}[pc] \gets K_{\text{cov}}[pc] + 1$
            \State $A \gets A + pc$
            \State \textbf{continue}
        \ElsIf{$pc \in F_{\text{cov}}$}
            \State $F_{\text{cov}}[pc] \gets F_{\text{cov}}[pc] + 1$
            \State $A \gets A + pc$
            \State \textbf{continue}
        \EndIf
        \If{\textsc{AddrInFilters}($pc, K_{\text{filter}}, K_{\text{starts}}$)}
            \State $K_{\text{cov}}[pc] \gets 1$
            \State $K_{\text{found}} \gets \textbf{true}$
            \State $A \gets A + pc$
            \State \textbf{continue}
        \EndIf
        \If{\textsc{AddrInFilters}($pc, F_{\text{filter}}, F_{\text{starts}}$)}
            \State $F_{\text{cov}}[pc] \gets 1$
            \State $F_{\text{found}} \gets \textbf{true}$
            \State $A \gets A + pc$
            \State \textbf{continue}
        \EndIf
    \EndFor
\end{algorithmic}
\end{algorithm}

\FloatBarrier

\vskip.5\baselineskip
\noindent\textbf{Snapshot:}

\textit{MTCFuzz} uses QEMU’s snapshot functionality to save and load status of virtual machine. These operations are performed through the QMP API. Snapshot saving through QMP is performed using the \texttt{snapshot-save} command, 
which stores the snapshot data on a storage device attached to the virtual machine. 
Consequently, as part of the setup process, \textit{MTCFuzz} creates a dedicated qcow2 file to store snapshots and launches QEMU with snapshot storage attached by the command-line option shown in \autoref{code:qemu-snapshot-drive}. 
In this configuration, the storage device is assigned the ID \texttt{snapshot0}. This ID is required when saving snapshots using the \texttt{snapshot-save} API. 
The general procedure for saving a snapshot is shown in \autoref{algo:savevm}. First, \textit{MTCFuzz} establishes a connection to QMP using \textsc{ConnectQMP}. 
Then it checks whether the \texttt{node\_name} of the snapshot storage is already set; if not, the node is searched using \texttt{FindBlockDevice}. 
Once the node is found, the execution of the virtual machine is suspended using \textsc{StopVM}. The snapshot target devices are then configured, the \texttt{snapshot-save} API parameters are prepared, 
and the \texttt{snapshot-save} command is executed. 
After the snapshot has been successfully saved, the virtual machine execution is resumed and the connection to QMP is closed, completing the procedure.

The procedure for loading a snapshot requires the \texttt{node\_name} of the storage device where the snapshot is stored. 
If the node name is not already set, it is obtained using \textsc{FindBlockDevice}, in the same manner as in the save operation. 
The virtual machine execution is then suspended using \textsc{StopVM}, the parameters for the \texttt{snapshot-load} API are configured, and the \texttt{snapshot-load} command is executed. 
Once the snapshot loading is complete, the execution of the virtual machine is resumed, and the connection to QMP is closed, completing the operation.

\begin{lstlisting}[
    language=bash,
    float=th,
    caption={QEMU command-line option for snapshot feature},
    label={code:qemu-snapshot-drive},
    breaklines=true,
    basicstyle=\ttfamily\normalsize
]
-drive file=/home/build/projects/mtcfuzz/work/fuzz-snapshot.qcow2,if=none,format=qcow2,id=snapshot0
\end{lstlisting}

\begin{algorithm}[th]
\centering
\caption{VM Snapshot Save via QMP}
\label{algo:savevm}
\begin{algorithmic}[1]
    \Require QMP socket path, snapshot device name
    \Ensure Snapshot is saved successfully
    \If{\textsc{ConnectQMP}() == \textbf{false}}
        \State return false
    \EndIf
    \If{$node\_name$ is not set}
        \State $node\_name \gets$ \textsc{FindBlockDevice}()
    \EndIf
    \If{$node\_name$ is \textbf{null}}
        \State \textsc{DisconnectQMP}()
        \State return false
    \EndIf
    \State \textsc{StopVM}()
    \State $devices \gets [node\_name]$
    \If{$rootfs\_device\_name$ exists}
        \State $devices \gets devices \cup \{rootfs\_device\_name\}$
    \EndIf
    \State $args \gets$ \{
        job-id: \textsc{GenerateSnapshotJobID}("save"),\\
        \hspace{1.8em}tag: ``mtcfuzz-snapshot'',\\
        \hspace{1.8em}vmstate: $node\_name$,\\
        \hspace{1.8em}devices: $devices$
    \}
    \State \textsc{QMPExecute}(``snapshot-save'', $args$)
    \State \textsc{ResumeVM}()
    \State \textsc{DisconnectQMP}()
    \State return true
\end{algorithmic}
\end{algorithm}

\begin{algorithm}[th]
\centering
\caption{VM Snapshot Load via QMP}
\label{algo:loadvm}
\begin{algorithmic}[1]
    \Require QMP socket path, snapshot tag
    \Ensure VM state is restored from snapshot
    \If{\textsc{ConnectQMP}() == \textbf{false}}
        \State return false
    \EndIf
    \If{$node\_name$ is not set}
        \State $node\_name \gets$ \textsc{FindBlockDevice}()
    \EndIf
    \State \textsc{StopVM}()
    \State $args \gets$ \{
        job-id: \textsc{GenerateSnapshotJobID}("load"),\\
        \hspace{1.8em}tag: ``mtcfuzz-snapshot'',\\
        \hspace{1.8em}vmstate: $node\_name$,\\
        \hspace{1.8em}devices: $[node\_name]$
    \}
    \State \textsc{QMPExecute}(``snapshot-load'', $args$)
    \State \textsc{ResumeVM}()
    \State \textsc{DisconnectQMP}()
    \State return true
\end{algorithmic}
\end{algorithm}

\FloatBarrier

\vskip.5\baselineskip
\noindent\textbf{Fuzzing workflow in MTCFuzz:}

\begin{figure}[tb]
\centering
\includegraphics[width=\hsize]{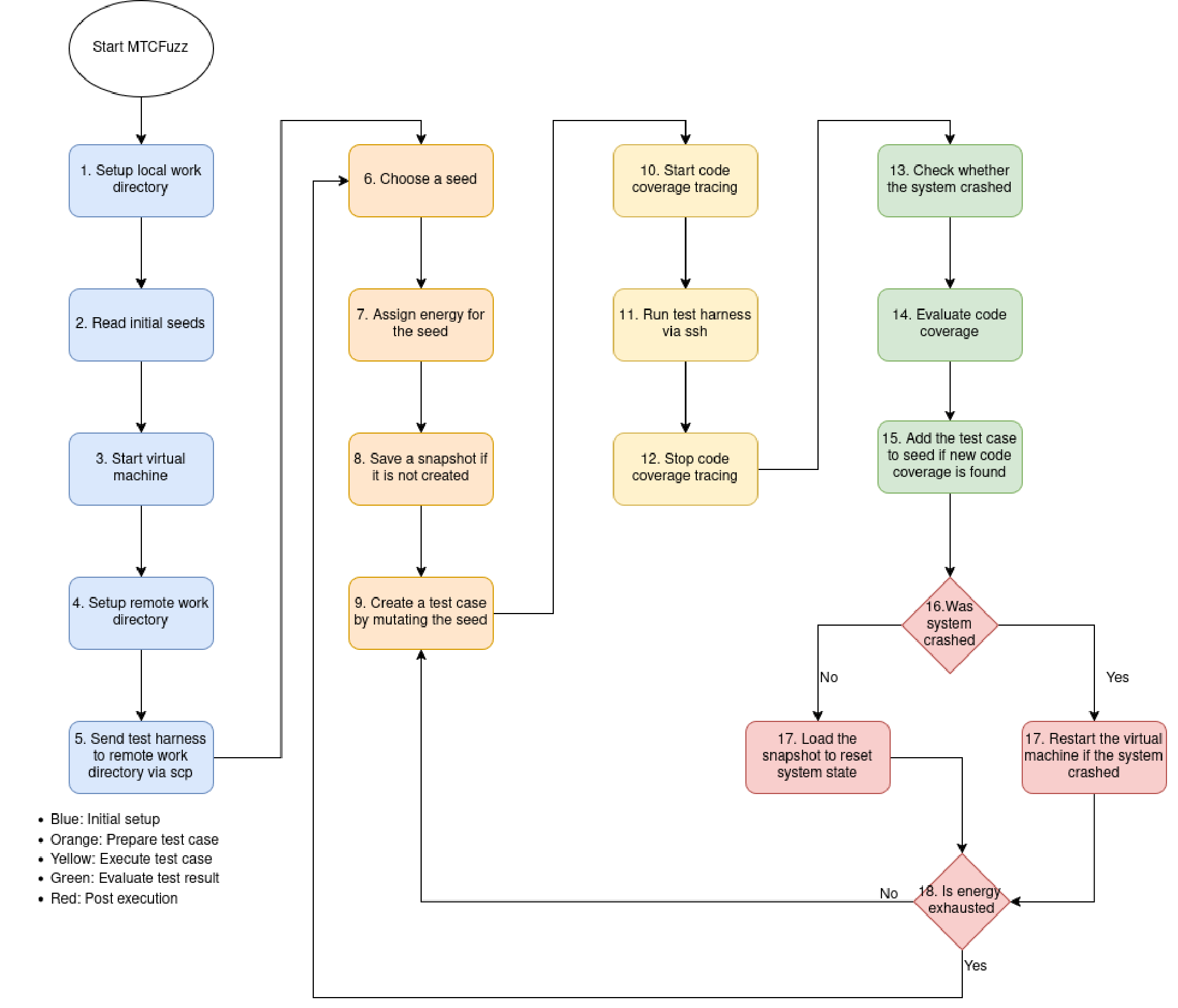}
\caption{Overview of the fuzzing workflow in MTCFuzz.}
\label{fig:fuzzing-workflow}
\end{figure}

\autoref{fig:fuzzing-workflow} illustrates the basic fuzzing execution workflow
of \textit{MTCFuzz}.
The workflow can be roughly divided into five phases:

\begin{enumerate}
  \item initialization,
  \item test case preparation,
  \item test case execution,
  \item evaluation of test execution results, and
  \item post-execution processing.
\end{enumerate}

The blue rectangles in \autoref{fig:fuzzing-workflow} represent the initialization phase.
This phase includes one-time setup operations performed before starting fuzzing,
such as creating a working directory on the host running \textit{MTCFuzz},
loading the initial seed corpus, launching the virtual machine,
creating a working directory inside the virtual machine,
and transferring the test harness to the virtual machine.

The orange rectangles correspond to the test case preparation phase,
which includes selecting a seed, assigning energy to the selected seed,
creating a snapshot if it has not been created yet,
and generating a test case by mutating the seed.
The yellow rectangles represent the test case execution phase,
which consists of starting code coverage tracing,
executing the test harness, and stopping code coverage tracing.
The green rectangles indicate the evaluation phase,
in which code coverage is evaluated based on the execution results.

The red rectangles represent the post-execution processing phase.
If the system crashes, the virtual machine is restarted;
otherwise, the system state is restored to the pre-execution state
by loading the snapshot.
When the energy assigned to the current seed is exhausted,
a new seed is selected; otherwise, mutation of the current seed is continued.
\textit{MTCFuzz} advances fuzzing by repeatedly executing this workflow.

Test cases generated by \textit{MTCFuzz} on the host side are either transferred
to the virtual machine as files via the \texttt{scp} command
or directly supplied as arguments to the test harness.
Since the test harness is executed inside the virtual machine,
it is invoked via remote command execution over \texttt{ssh},
as shown in \autoref{code:execute-test-case-via-ssh}.






\begin{lstlisting}[
    language=bash,
    float=th,
    caption={Execute test case via ssh},
    label={code:execute-test-case-via-ssh},
    breaklines=true,
    basicstyle=\ttfamily\normalsize
]
ssh -p 10022 localhost /path/to/testharness arg1 arg2 ...
\end{lstlisting}

\FloatBarrier

\vskip.5\baselineskip
\noindent\textbf{Visualizing code coverage:}

\begin{figure}[tb]
\centering
\includegraphics[width=\hsize]{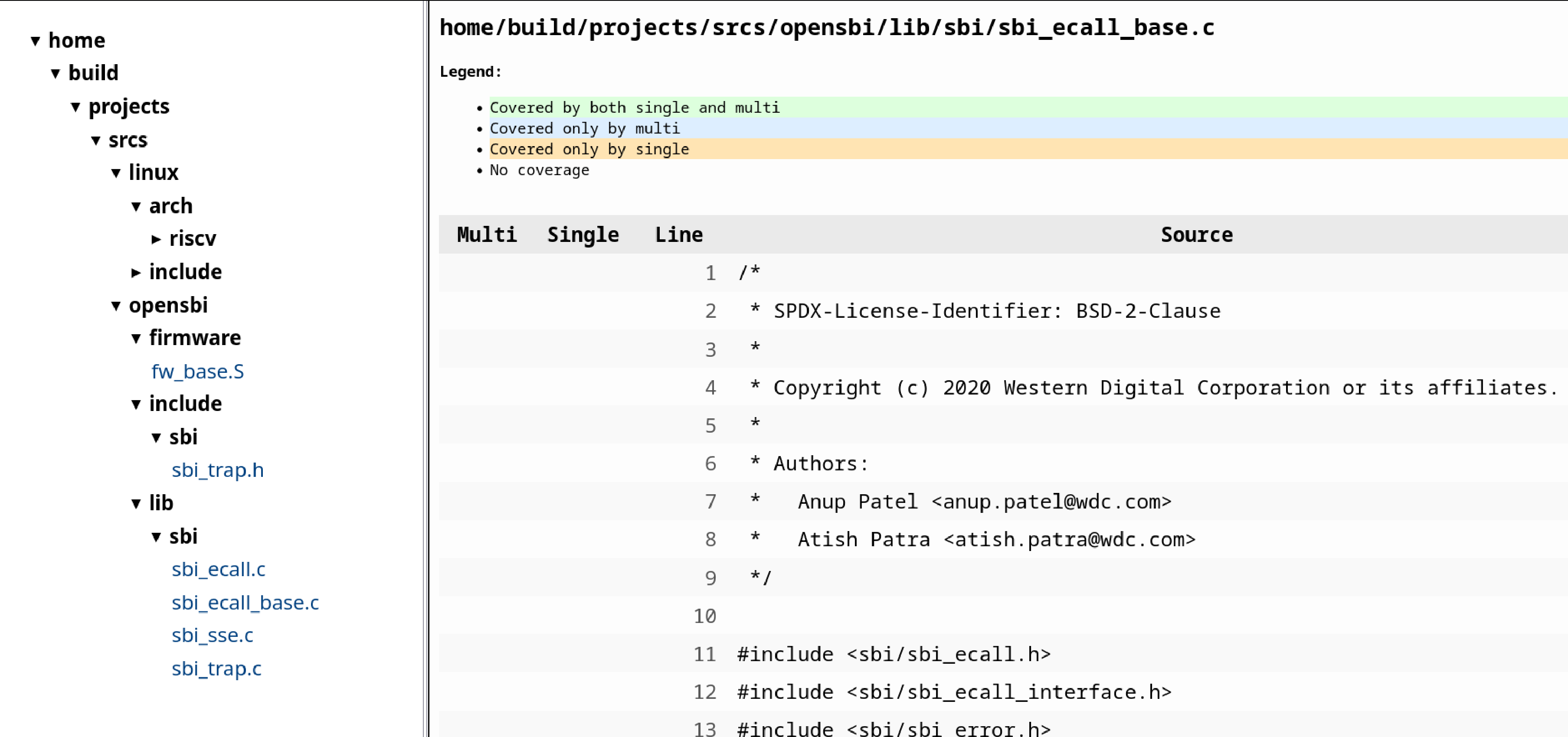}
\caption[
Coverage visualization in HTML
]{
Coverage visualization in HTML, illustrating the meaning of background colors.
}
\label{fig:coverage-html-top}
\end{figure}

\begin{figure}[tb]
\centering
\includegraphics[width=\hsize]{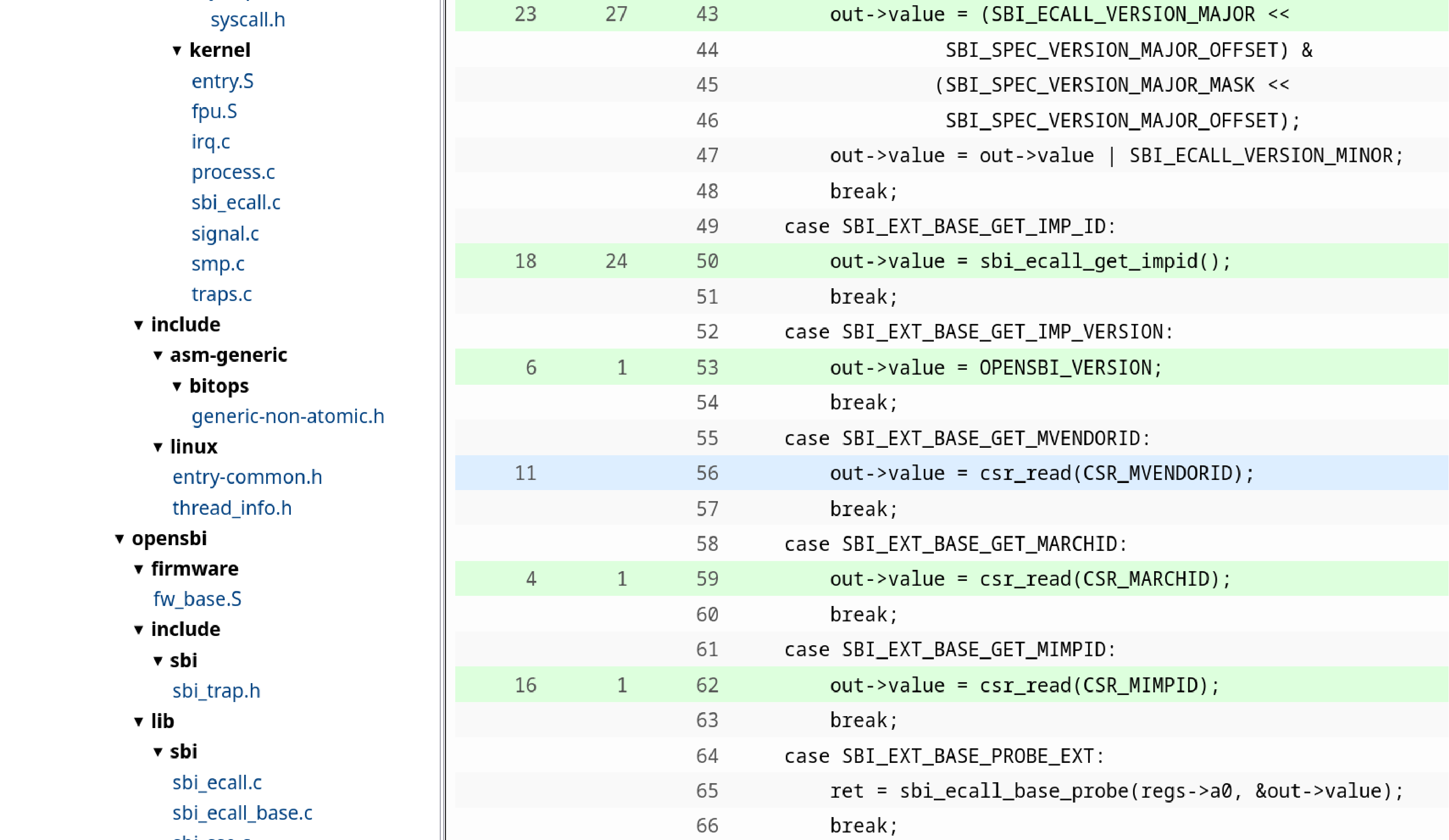}
\caption[
Coverage differences between \texttt{multi} and \texttt{single}
]{
Illustrating coverage differences between the proposed method (\texttt{multi}) and the baseline (\texttt{single}).
The blue-highlighted line at line 56 was executed only in the multi configuration.
}
\label{fig:coverage-visualize}
\end{figure}

\textit{MTCFuzz} provides a feature that visualizes the code coverage executed after fuzzing in an HTML format. 
It enables users to correlate executed coverage with the corresponding source code. This feature allows users to inspect source code regions that were executed during fuzzing.
Although this functionality is primarily developed to visually compare the results of the proposed method with those of a baseline approach, 
it can also be used for post-fuzzing inspection in standard fuzzing workflows that do not involve such comparisons. 
Since \textit{MTCFuzz} measures coverage at the basic block granularity, the correspondence between source code lines and coverage information is not strictly one-to-one. Nevertheless, 
the ability to identify executed regions is helpful for debugging purposes. In the HTML view, executed lines are highlighted with background colors: lines executed by both the proposed method and the baseline approach are shown in green, lines executed only by the proposed method are shown in blue, and lines executed only by the baseline approach are shown in orange. 
As illustrated in \autoref{fig:coverage-html-top}, selecting a source file displays a legend explaining these background colors at the top of the page. 
An example of visualizing executed regions within a selected source file is shown in \autoref{fig:coverage-visualize}, 
where the number of executions is displayed for each executed line. 
In the \autoref{fig:coverage-visualize}, the blue-highlighted line at line 56 indicates that it was executed twice only by the proposed method.

This functionality correlates executed basic blocks with source code and the target software to be built with debug information enabled. 
As a result, this feature cannot be used when fuzzing closed-source software for which source code is unavailable. 
To resolve executed basic blocks to source code locations, we uses the \texttt{addr2line} command. 
The \texttt{addr2line} tool can be applied to binaries in the \texttt{ELF} file format, as shown in \autoref{code:addr2line-example}. 
In addition to specifying target addresses directly, \texttt{addr2line} supports batch processing by reading multiple addresses from a file. In this example, 
\texttt{addr2line} outputs the function name, source file path, and line number corresponding to each address. 
In \textit{MTCFuzz}, the addresses to be resolved are written to a file and processed in a single invocation of the \texttt{addr2line} command to obtain this information efficiently. Although \texttt{addr2line} operates on binaries in the \texttt{ELF} file format, firmware such as \texttt{OpenSBI} is typically deployed as a raw binary rather than 
an \texttt{ELF} file. However, for debugging purposes, such firmware projects also generate binaries in the \texttt{ELF} format. 
Therefore, when invoking \texttt{addr2line}, \textit{MTCFuzz} specifies the corresponding \texttt{ELF} file instead of the raw binary that is actually loaded at runtime.

\begin{lstlisting}[
language=bash,
caption={Resolving a kernel address using addr2line},
label={code:addr2line-example},
breaklines=true,
basicstyle=\ttfamily\normalsize
]
$ riscv64-linux-gnu-addr2line -e ./vmlinux -f -p -a 0xffffffff8001b340
0xffffffff8001b340: __sbi_ecall at /home/build/projects/srcs/linux/arch/riscv/kernel/sbi_ecall.c:41
\end{lstlisting}

\FloatBarrier

\section{Summary of the Proposed Approach}

In this chapter, we proposed the \textit{Multi-target Coverage-based Greybox Fuzzing} approach
that integrates and utilizes the code coverage of multiple software components cooperating
within a system. Conventional CGF primarily targets a single piece of software; therefore,
in systems where software cooperates with other components, such as the Linux kernel and
OpenSBI, it is difficult to observe the behavior of the cooperating software components.
In addition, Compiler-based coverage instrumentation provided by GCC or Clang is only
applicable to software that is built with such instrumentation enabled, and in systems
where multiple software components operate cooperatively, it is difficult to measure code
coverage continuously in temporal order.

In our proposed method, QEMU is adopted as the execution environment, and the code
coverage of software running on QEMU is measured within QEMU. It enables unified
coverage measurement regardless of how each software component is built. Furthermore,
by extending QMP to provide measurement control functionality and by utilizing the
snapshot mechanism to reset the execution environment, the influence of previous fuzzing
runs on the system state can be eliminated, thereby enabling a practical fuzzing
environment.

In the proposed method, the regions for which code coverage should be measured can be
specified, enabling the address filtering mechanism that excludes executions unrelated to
the target functionality from the coverage measurement. 
The computational complexity of the filtering process in the proposed method is $O(1)$ in the best case and $O(M log N)$ in the worst case, while computational complexity towards $O(1)$ on average. 
Therefore, even if the number of registered filters increases, its impact on filtering performance remains small.

Based on these designs, we implemented a prototype system called \textit{MTCFuzz} and
constructed a platform capable of evaluating the effectiveness of the proposed method on
multiple CPU architectures. In the next chapter, we evaluate the effectiveness of the proposed
method through experiments and present the results.
In the next chapter, we evaluate how effectively \textit{MTCFuzz} explores cooperative system software in practice.

\chapter{Evaluation}
\label{ch:Evalation}
In this chapter, we describe evaluation environments, evaluation methods, and evaluation results.

\begin{table}[tb]
\centering
\caption{Experimental environments}
\label{tab:experimental-environments}
\begin{tabular}{lll}
\hline
\textbf{Environment} & \textbf{OS} & \textbf{Hardware} \\
\hline
Env1 & Ubuntu 24.04 &
CPU: Intel Core i7-9700K \\
      &                &
Memory: 64\,GB \\[2mm]
Env2 & Fedora 43 &
CPU: Intel Core i7-14700 \\ 
      &              &
Memory: 64\,GB \\
\hline
\end{tabular}
\end{table}

\begin{table}[tb]
\centering
\caption{Software versions used in the experiments}
\label{tab:software-versions}
\begin{tabular}{lll}
\hline
\textbf{Environment} & \textbf{Component} & \textbf{Version} \\
\hline
\multirow{2}{*}{RISC-V} 
  & Linux Kernel & 6.16-rc1 \\
  & OpenSBI      & v1.6 \\
\hline
\multirow{3}{*}{OP-TEE} 
  & Manifest     & \texttt{f92aacd1b103af137438f2e434303ffa5dff3d09} \\
  & Linux Kernel & 6.14.0-gb8a233c2155a \\
  & OP-TEE       & 4.7.0-rc1 \\
\hline
\end{tabular}
\end{table}

\section{Experimental Environment}

The experiments were conducted using the two environments shown in \autoref{tab:experimental-environments}. The experiments were conducted on two architectures: RISC-V and AArch64. The software used for these architectures is listed in \autoref{tab:software-versions}. The fuzzing experiments were performed using the Docker container provided by \textit{MTCFuzz}.

\section{Experimental Methodology}

Our proposed method improves the capability of CGF by unifying and using the code coverage obtained from multiple software components. For the experimental evaluation, 
\textit{MTCFuzz} provides two execution modes: a mode that utilizes the coverage of multiple software components (which realizes the proposed method) and a mode that utilizes the code coverage of only a single software component (as in conventional fuzzers).
We refer to execution with the proposed method as \texttt{multi}, and execution using only single software code coverage as \texttt{single}.  In this study, the \texttt{single} mode uses only the Linux kernel code coverage. There is virtually no difference in the execution code between these two modes. In both cases, code coverage is collected; however, in the \texttt{single} mode, 
the firmware side coverage is always treated as if no new coverage has been discovered. 
As a result, the number of steps required for code coverage measurement is identical in both modes, and no performance difference arises from the coverage measurement process itself. 
In our experimental evaluation of the proposed method, code coverage effectiveness is assessed by comparing the results of \texttt{multi} and \texttt{single} using this functionality.

In all experiments, the number of QEMU instances launched by \textit{MTCFuzz} was set to one.  
The configuration files used for each fuzzing experiment are summarized in \autoref{tab:fuzzing-configs}.

\begin{table}[tb]
\centering
\caption{Configuration files used in fuzzing experiments}
\label{tab:fuzzing-configs}
\small
\resizebox{\textwidth}{!}{%
\begin{tabular}{p{0.55\linewidth} p{0.35\linewidth}}
\toprule
Experiment & Configuration file \\
\midrule
\ref{sec:OpenSBI-Base-Extension-coverage}\\
: \nameref{sec:OpenSBI-Base-Extension-coverage}
& \texttt{configs/opensbi/coverage\_test} \\

\ref{sec:CGF-with-Multi-Target-Coverage-Feedback}\\
: \nameref{sec:CGF-with-Multi-Target-Coverage-Feedback}
& \texttt{configs/optee/xtest\_fuzz\_1002} \\

\ref{sec:Address-filter-performance}\\
: \nameref{sec:Address-filter-performance}
& \texttt{configs/opensbi/coverage\_test} \\

\ref{sec:snapshot-load-performance}\\
: \nameref{sec:snapshot-load-performance}
& \texttt{configs/opensbi/coverage\_test} \\

\ref{sec:cve-finding}\\
: \nameref{sec:cve-finding}
& \texttt{configs/optee/xtest\_fuzz\_1001} \\

\bottomrule
\end{tabular}}
\end{table}

\FloatBarrier

\subsection{Coverage Evaluation of OpenSBI Base Extension}
\label{sec:OpenSBI-Base-Extension-coverage}

In this evaluation, we used the RISC-V architecture with Linux kernel 6.16-rc1 and OpenSBI v1.6 in \texttt{Env2}. The evaluation focuses on measuring and analyzing the coverage of \textsc{case} 
statements within the \texttt{sbi\_ecall\_base\_handler} function, which handles the \texttt{Base Extension} in \texttt{lib/sbi/sbi\_ecall\_base.c} of OpenSBI. 
The \texttt{Base Extension} provides fundamental APIs, such as those that return the implemented specification version. According to the RISC-V Supervisor Binary Interface Specification, 
the \texttt{Base Extension} defines seven APIs~\cite{riscv_sbi_2_0}. In the OpenSBI implementation, the extension ID is stored in the \texttt{eid} variable, and the function ID is stored in the \texttt{fid} variable. 
The \texttt{Base Extension} has a value of \texttt{0x10}, and \texttt{sbi\_ecall\_base\_handler} executes the corresponding process according to the \texttt{fid} value. To perform branching based on the value of \texttt{fid}, 
this implementation uses a \texttt{switch} statement.
In this evaluation, the mutation target was set to \texttt{fid} and the initial seed used a value of 0 for it. 
Based on this initial seed, the fuzzing was conducted in two independent sets, each consisting of 50 campaigns of 5 minutes, resulting in a total of 100 fuzzing campaigns.

\subsection{Coverage evaluation of OP-TEE by long running}
\label{sec:CGF-with-Multi-Target-Coverage-Feedback}
In this experiment, we compare the number of covered code paths between \texttt{multi} and \texttt{single} during fuzzing of the shared memory functionality used jointly by the Linux kernel and the Trusted OS.

\subsection{QEMU performance overhead}

For QEMU performance evaluation, we used a RISC-V Linux environment running kernel version 6.16-rc1 as the guest OS. The number of virtual CPUs in the guest environment was set to one. 
To measure performance, we used \texttt{hackbench}~\cite{hackbench}, which is included in the Ubuntu 24.04 rt-tests package. 
The \texttt{hackbench} is a benchmark and stress testing tool for the Linux kernel scheduler. To run a stress test by \texttt{hackbench}, it does not use storage. Therefore, test result is unaffected by I/O operations. 
This allows us to measure kernel processing performance in a purely CPU- and scheduler-bound manner. 
We evaluate two configurations of \texttt{hackbench}: the \textbf{process mode}(\texttt{hackbench -l 600 -g 10 --process}) and the \textbf{thread mode}(\texttt{hackbench -l 600 -g 10 --thread}). 
Both configurations were executed with 600 loops and 10 groups. As a baseline for comparison, we used the same version of QEMU(v9.2.3) built and used by \textit{MTCFuzz}.

\subsection{Address filter performance}
\label{sec:Address-filter-performance}

For the evaluation of the address filter, the same fuzzing test described in \ref{sec:OpenSBI-Base-Extension-coverage} is used. The number of configured filters increases from $2^{1}$ to $2^{15}$, and a fuzzing campaign was run for 10 minutes for each test. The upper limit of the number of filters was set to $2^{15}$. As shown in \autoref{code:risv-linux-kallsyms}, there are 51,779 symbols in the tested kernel.
Also, as shown in \autoref{code:f43-kallsyms}, the number of symbols in Fedora~43, a widely used Linux distribution (Linux kernel version 6.17.8-300.fc43.x86\_64), is 403{,}481. It is corresponds to $ \log_{2}(403{,}481) \approx 18.62$, and therefore we consider that setting the maximum number of filters to $2^{15}$ is sufficient for the purpose of this experiment.

\begin{lstlisting}[
    language=c, 
    float=tb, 
    caption={Fedora 43 total symbols},
    label={code:f43-kallsyms},
    numbers=left,
    basicstyle=\ttfamily\normalsize,
    breaklines=true
]
$ uname -r
6.17.8-300.fc43.x86_64
$ sudo wc -l /proc/kallsyms 
403481 /proc/kallsyms
\end{lstlisting}

\begin{lstlisting}[
    language=c, 
    float=tb, 
    caption={Total symbols in tested RISC-V kernel},
    label={code:risv-linux-kallsyms},
    numbers=left,
    basicstyle=\ttfamily\normalsize,
    breaklines=true
]
# uname -a
Linux buildroot 6.16.0-rc1 #2 SMP Sun Aug 24 03:10:39 UTC 2025 riscv64 GNU/Linux
# wc -l /proc/kallsyms 
51779 /proc/kallsyms
\end{lstlisting}

\FloatBarrier

\subsection{Snapshot load performance}
\label{sec:snapshot-load-performance}

\textit{MTCFuzz} loads a snapshot before the execution of each test case and restores the environment to its pre-execution state. This mechanism enables each test to be executed without being affected by the results of previous tests and is used to minimize the occurrence of bugs that depend on runtime state. However, restoring the snapshot for every test case may introduce performance overhead during fuzzing.

Therefore, in the experiment, we compared the number of test executions between two configurations: (i) restoring the snapshot before each test case (the default behavior of \textit{MTCFuzz}), and (ii) continuing fuzzing without restoring the snapshot. The experiment uses the same configuration as in \ref{sec:OpenSBI-Base-Extension-coverage} (\nameref{sec:OpenSBI-Base-Extension-coverage}), and both configurations—normal execution of \textit{MTCFuzz} and the execution with snapshot loading disabled—were evaluated. For each configuration, five times a set of 10-minute fuzzing runs was conducted and the results were compared.

\subsection{Fuzzing OP-TEE cryptographic API}
\label{sec:cve-finding}

In this experiment, we constructed a test harness that mutates parameters
associated with shared memory used in cryptographic operations in OP-TEE 
and conducted an evaluation based on this setup.
The target cryptographic operation accepts, as input parameters,
a buffer containing plain text data to be encrypted and its length,
and produces, as output parameters, a buffer for the resulting cipher text
and its length.
For APIs that take both a buffer and its corresponding size as arguments,
inconsistencies between the actual buffer size and the specified size may lead to
memory corruption.
Consequently, this evaluation aimed to detect vulnerabilities caused by
such buffer size mismatches.

\section{Experimental Results}

\subsection{Coverage Evaluation of OpenSBI Base Extension results}
\label{lab:Coverage-Evaluation-of-OpenSBI-Base-Extension-result}
In this experiment, a fuzzing campaign was set to 5 minutes and two sets of 50 executions were performed, resulting in a total of 100 fuzzing campaigns. \autoref{tab:full-coverage-count} shows the number of campaigns in which all branching points of the \texttt{switch} statement were fully covered in each test set and the total number of fully covered. The results indicate that \texttt{multi} achieved cover all branch points 27 times in Set 1, 28 campaigns in Set 2, then total 55 campaigns (55\%). In contrast, \texttt{single} achieves cover all branch points 19 times in Set 1, 15 campaigns in Set 2, then a total of 34 campaigns (34\%). \autoref{tab:avg-coverage} shows the average code coverage rate for each set, as well as the overall average code coverage rate. From the average results, \texttt{multi} achieved an improvement of approximately 5.9\% over \texttt{single} in terms of coverage. These experimental results demonstrate that the proposed method improves the code coverage rate by approximately 20\% compared to the conventional CGF that relies on the code coverage of a single software component. This indicates that the proposed method can achieve higher code coverage in a shorter period of time than the conventional approach.

\begin{table}[tb]
\centering
\caption{Number of campaigns that achieved full coverage}
\label{tab:full-coverage-count}
\begin{tabular}{lccc}
\hline
\textbf{Mode} & \textbf{Set 1} & \textbf{Set 2} & \textbf{Total} \\
\hline
\texttt{multi}  & 27 & 28 & 55 \\
\texttt{single} & 19 & 15 & 34 \\
\hline
\end{tabular}
\end{table}

\begin{table}[tb]
\centering
\caption{Average number of coverage campaigns}
\label{tab:avg-coverage}
\begin{tabular}{lccc}
\hline
\textbf{Mode} & \textbf{Set 1 Avg.} & \textbf{Set 2 Avg.} & \textbf{Overall Avg.} \\
\hline
\texttt{multi}  & 6.34 & 6.14 & 6.24 \\
\texttt{single} & 5.92 & 5.86 & 5.89 \\
\hline
\end{tabular}
\end{table}

\subsection{Coverage evaluation of OP-TEE by long running results}
\label{lab:CGF-with-Multi-Target-Coverage-Feedback-result}

\begin{figure}[!t]
\centering
\includegraphics[width=\hsize]{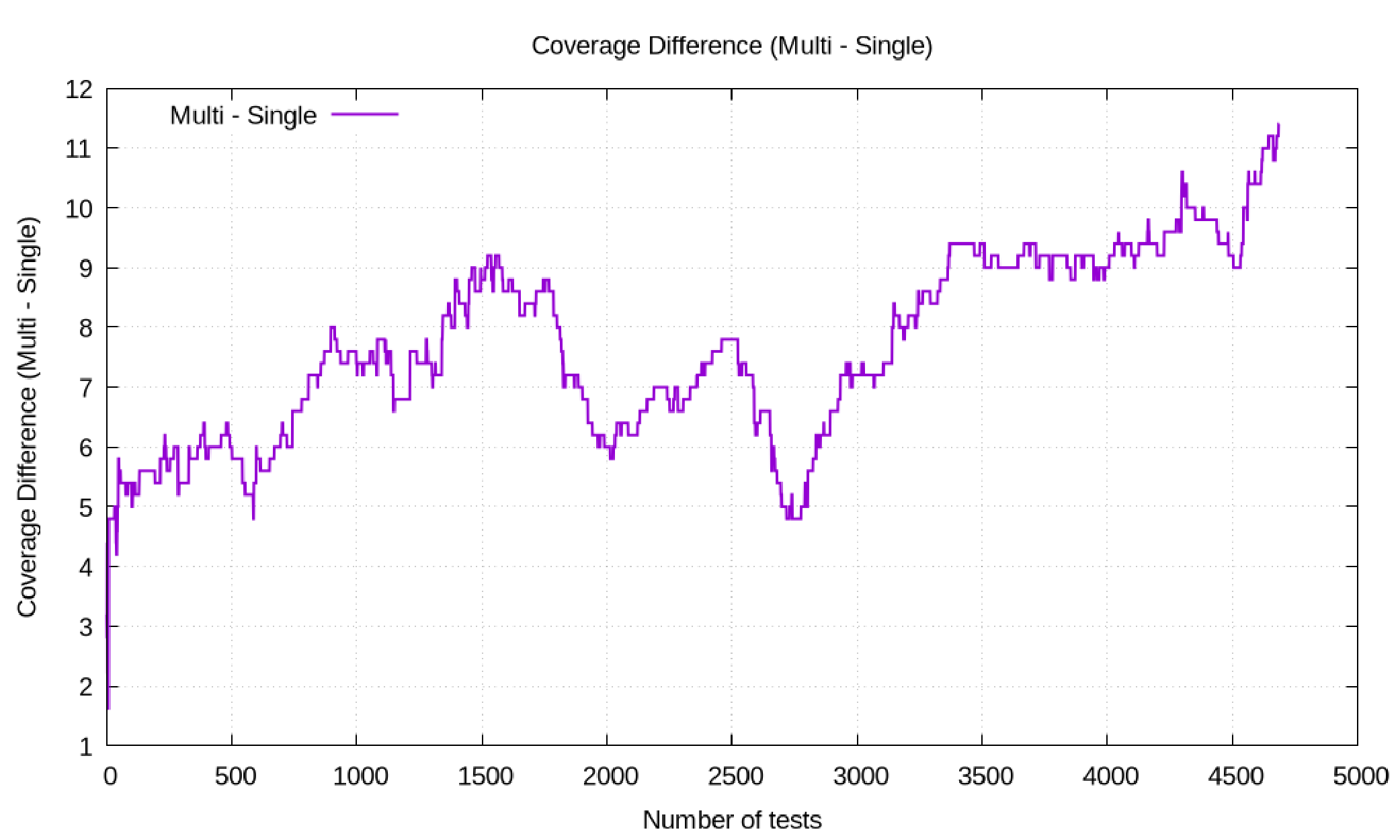}
\caption[
Coverage difference between Multi and Single (2 hours)
]{
Coverage difference (Multi - Single) between the Multi and Single configurations, aligned to 2 hours test executions.
}\label{fig:coverage-diff_2h}
\end{figure}

On \texttt{Env1}, we conducted fuzzing experiments for both \texttt{multi} and \texttt{single}, consisting of five campaigns with a duration of two hours each. Since the number of test executions varies across campaigns, we summarized the increase in code coverage for the first 4,688 test executions, which is a minimal test execution count at the time. For comparison, we computed the average code coverage for each test execution for both \texttt{multi} and \texttt{single}. \autoref{fig:coverage-diff_2h} visualizes the coverage difference between \texttt{multi} and \texttt{single} (\texttt{multi} $-$ \texttt{single}) for each test execution. In this graph, the zero value on the y-axis represents the origin where there is no difference in code coverage between \texttt{multi} and \texttt{single}. Positive values indicate that \texttt{multi} achieves higher code coverage, whereas negative values indicate that \texttt{single} achieves higher code coverage.

As shown in \autoref{fig:coverage-diff_2h}, in the comparison aligned to  4,688 test executions which are the same as above , the Multi configuration achieves higher coverage than the Single configuration at nearly all execution points, and the gap between them increases further in the latter half of the execution. These results indicate that, within a limited execution time budget, the \texttt{multi} provides higher exploration efficiency than the \texttt{single}.

\begin{figure}[!t]
\centering
\includegraphics[width=\hsize]{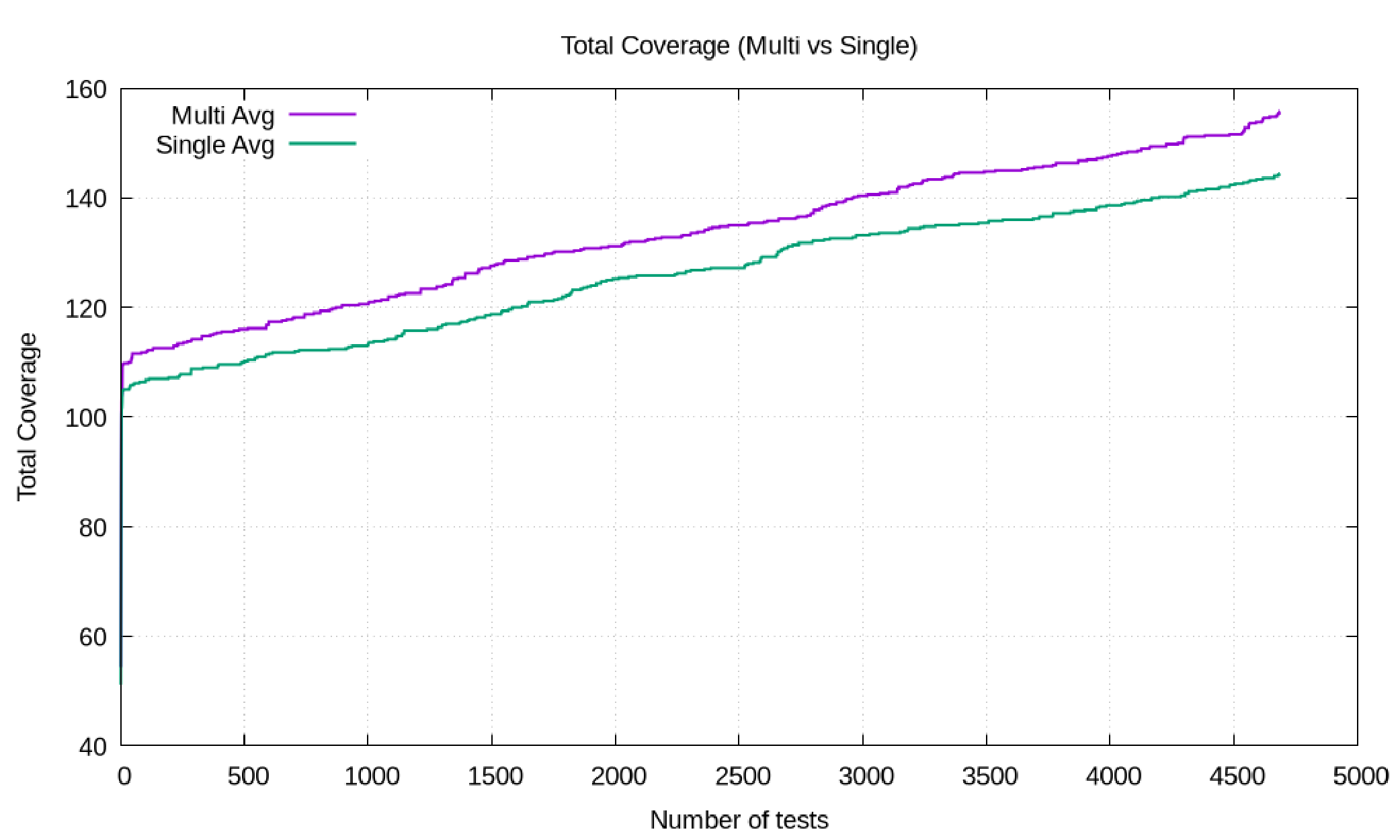}
\caption[
Total coverage of the Multi and Single (2 hours)
]{
Total coverage of the Multi and Single configurations, aligned to 2 hours of test executions.
}
\label{fig:coverage-multi-vs-single-2h}
\end{figure}

\begin{figure}[!t]
\centering
\includegraphics[width=\hsize]{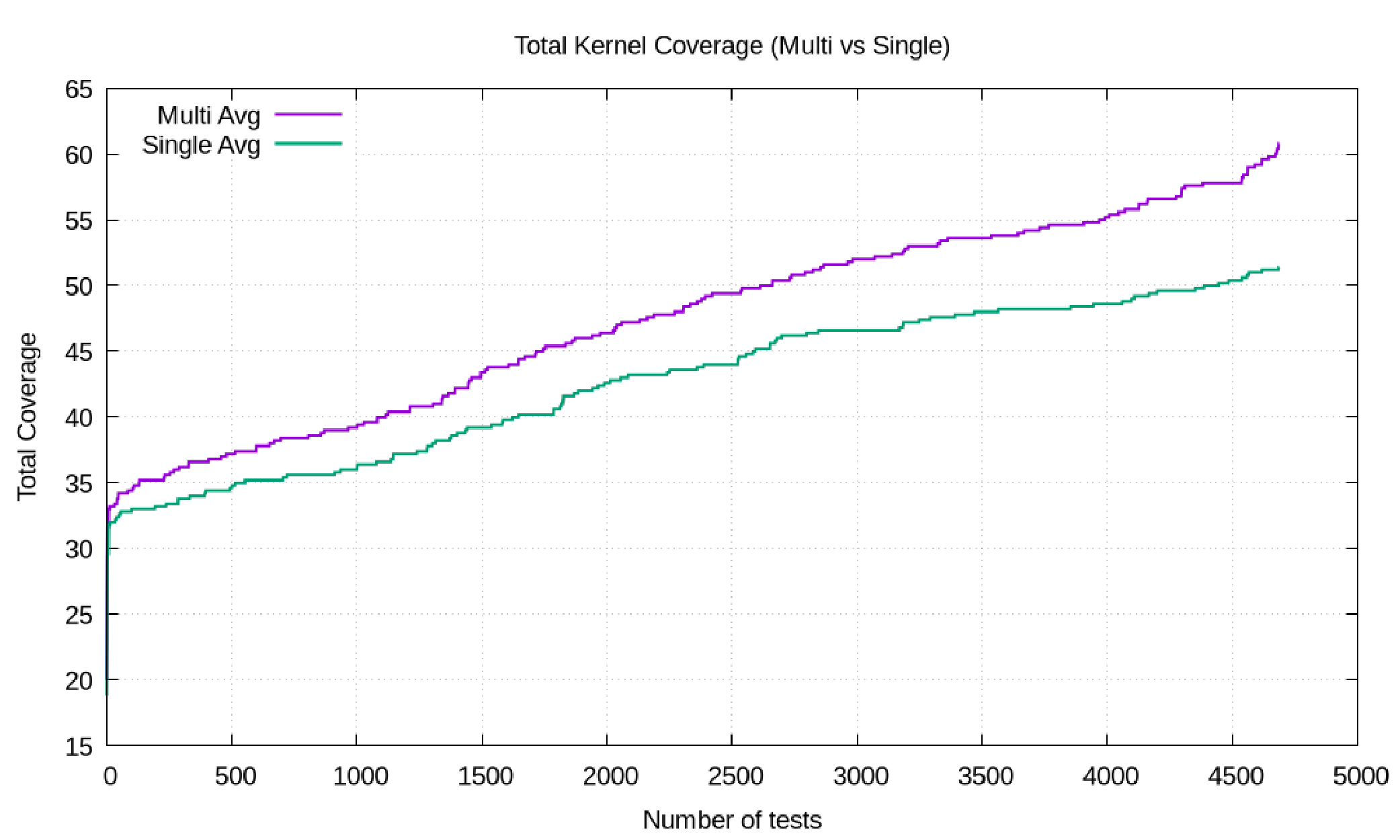}
\caption[
Total kernel coverage Multi vs. Single (2 hours)
]{
Total kernel coverage of the Multi and Single configurations, aligned to 2 hours of test executions.
}
\label{fig:coverage-multi-vs-single-2h-kernel}
\end{figure}

\begin{figure}[!t]
\centering
\includegraphics[width=\hsize]{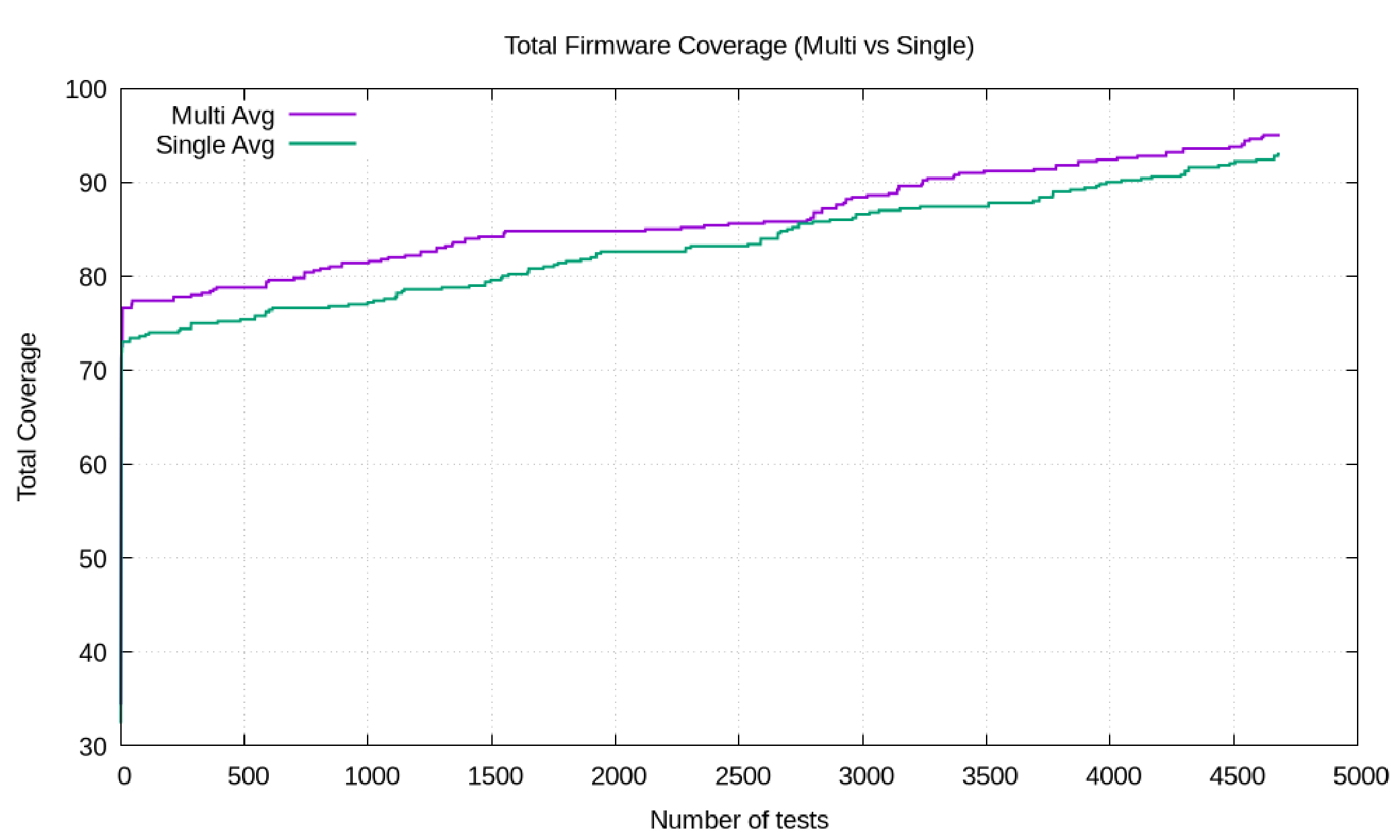}
\caption[
Total firmware coverage Multi vs. Single (2 hours)
]{
Total firmware coverage of the Multi and Single configurations, aligned to 2 hours of test executions.
}
\label{fig:coverage-multi-vs-single-2h-firmware}
\end{figure}

\autoref{fig:coverage-multi-vs-single-2h} illustrates the growth of code coverage. As the number of test executions increases, the gap between \texttt{multi} and \texttt{single} becomes larger. Furthermore, at nearly all execution points, \texttt{multi} consistently achieves higher code coverage than \texttt{single}. 
\autoref{fig:coverage-multi-vs-single-2h-kernel} and \autoref{fig:coverage-multi-vs-single-2h-firmware} show the growth of the code coverage separately for the kernel and the firmware. In the firmware code coverage, the coverage achieved by \texttt{single} catched up with that of \texttt{multi} around the 2,700th test execution; however, \texttt{multi} subsequently continued to discover additional coverage and ultimately succeeds in detecting more coverage than \texttt{single}.
From the experiment, we can conclude that the proposed method, \texttt{multi}, achieves code exploration efficiency on average higher than the conventional CGF targeting a single software component \texttt{single}.

\begin{figure}[!t]
\centering
\includegraphics[width=\hsize]{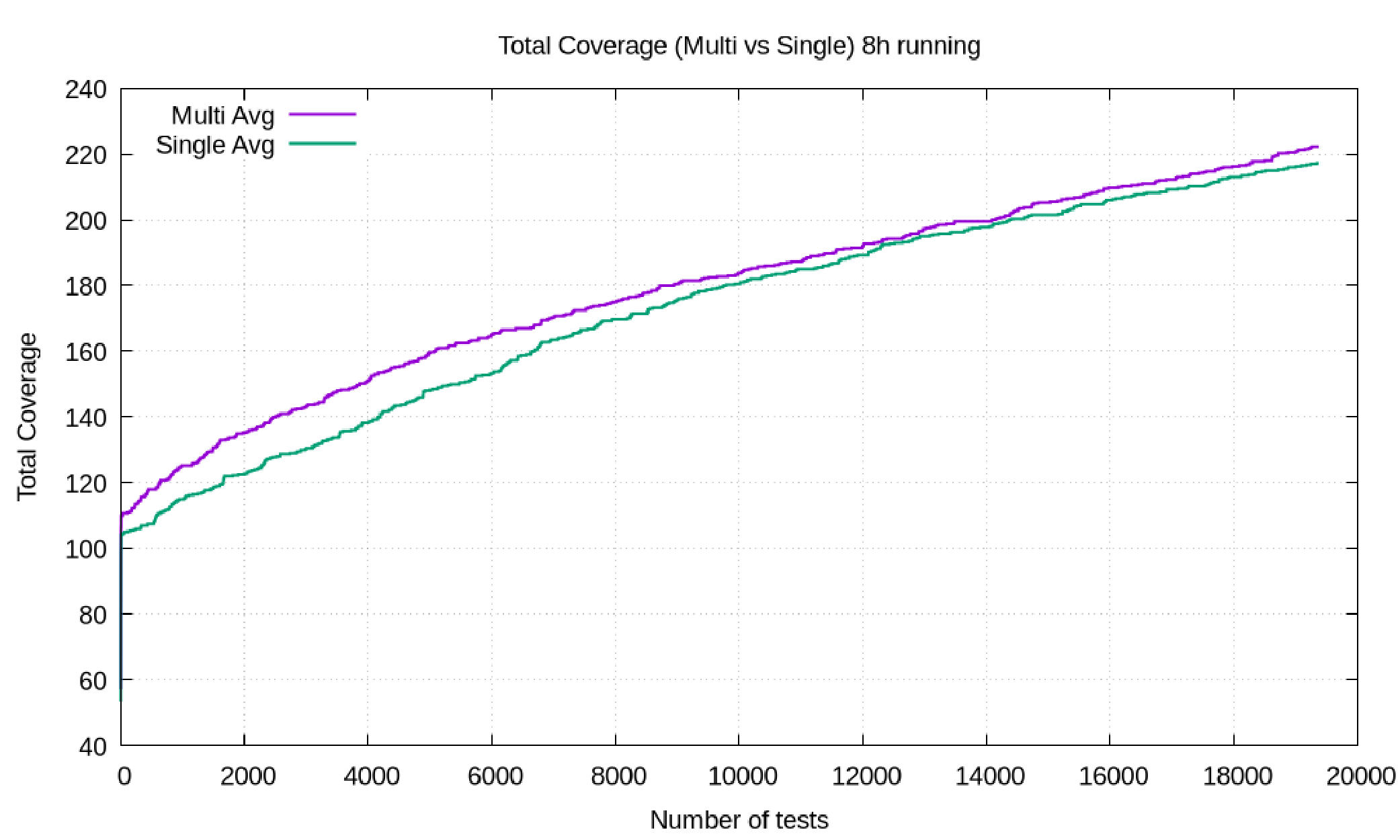}
\caption[
Total coverage Multi vs. Single (8 hours)
]{
Total coverage of the Multi and Single configurations, aligned to 8 hours of test executions.
}
\label{fig:coverage-multi-vs-single-8h}
\end{figure}

Here, the evaluation had been conducted using 2-hour fuzzing campaigns. In order to evaluate the behavior over a longer execution period, both \texttt{multi} and \texttt{single} fuzzing campaigns were configured to run for 8-hour fuzzing campaign, and each configuration was executed 4 times. The average of these runs was shown in \autoref{fig:coverage-multi-vs-single-8h}. 
In the figure, a relatively large difference between \texttt{multi} and \texttt{single} can be observed up to around 6000 test executions. Although this difference gradually decreases as the number of executed tests increases, \texttt{multi} continues to maintain higher coverage than \texttt{single}.

\begin{figure}[!t]
\centering
\includegraphics[width=\hsize]{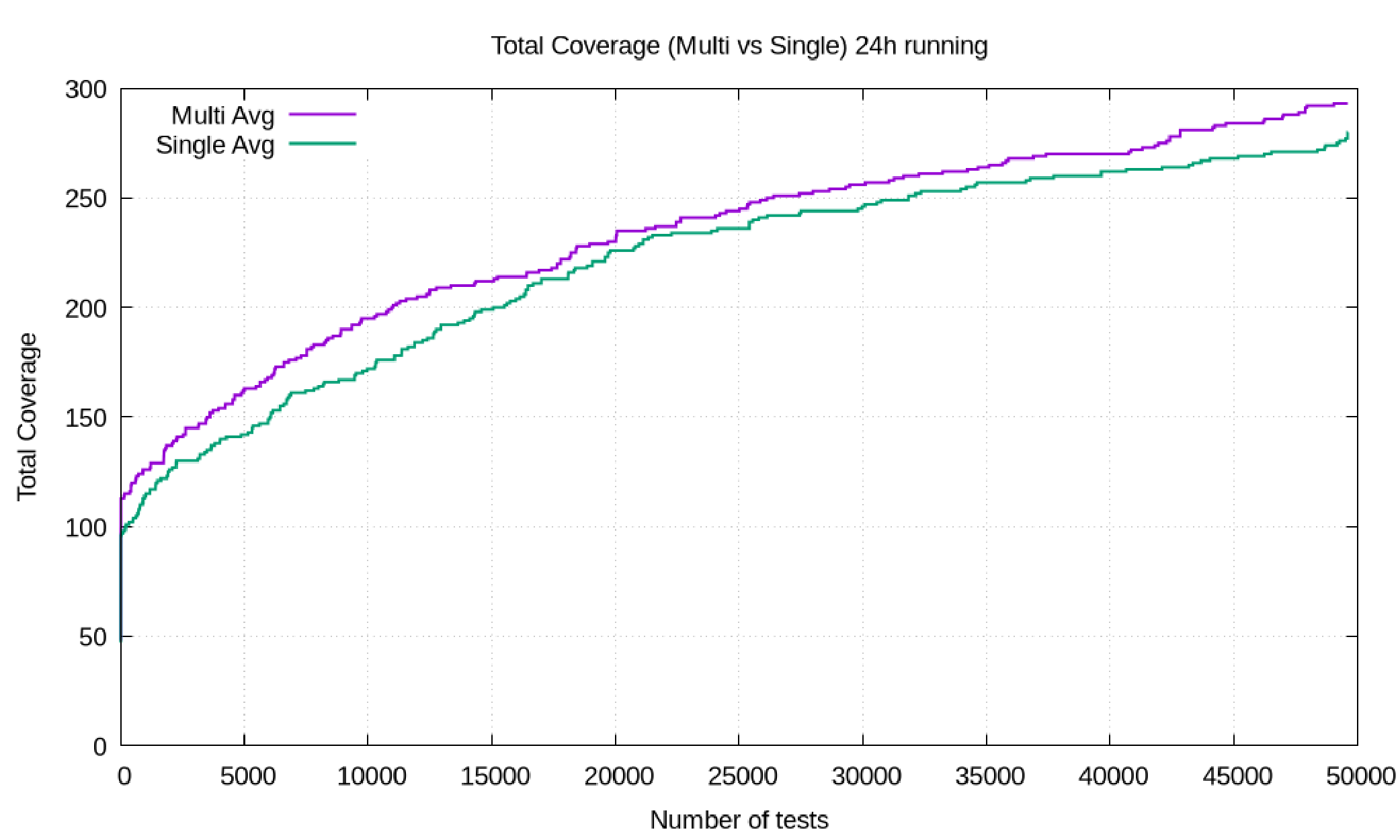}
\caption[
Total coverage Multi vs. Single (24 hours)
]{
Total coverage of the Multi and Single configurations, aligned to 24 hours of test executions.
}
\label{fig:coverage-multi-vs-single-24h}
\end{figure}

\begin{figure}[!t]
\centering
\includegraphics[width=\hsize]{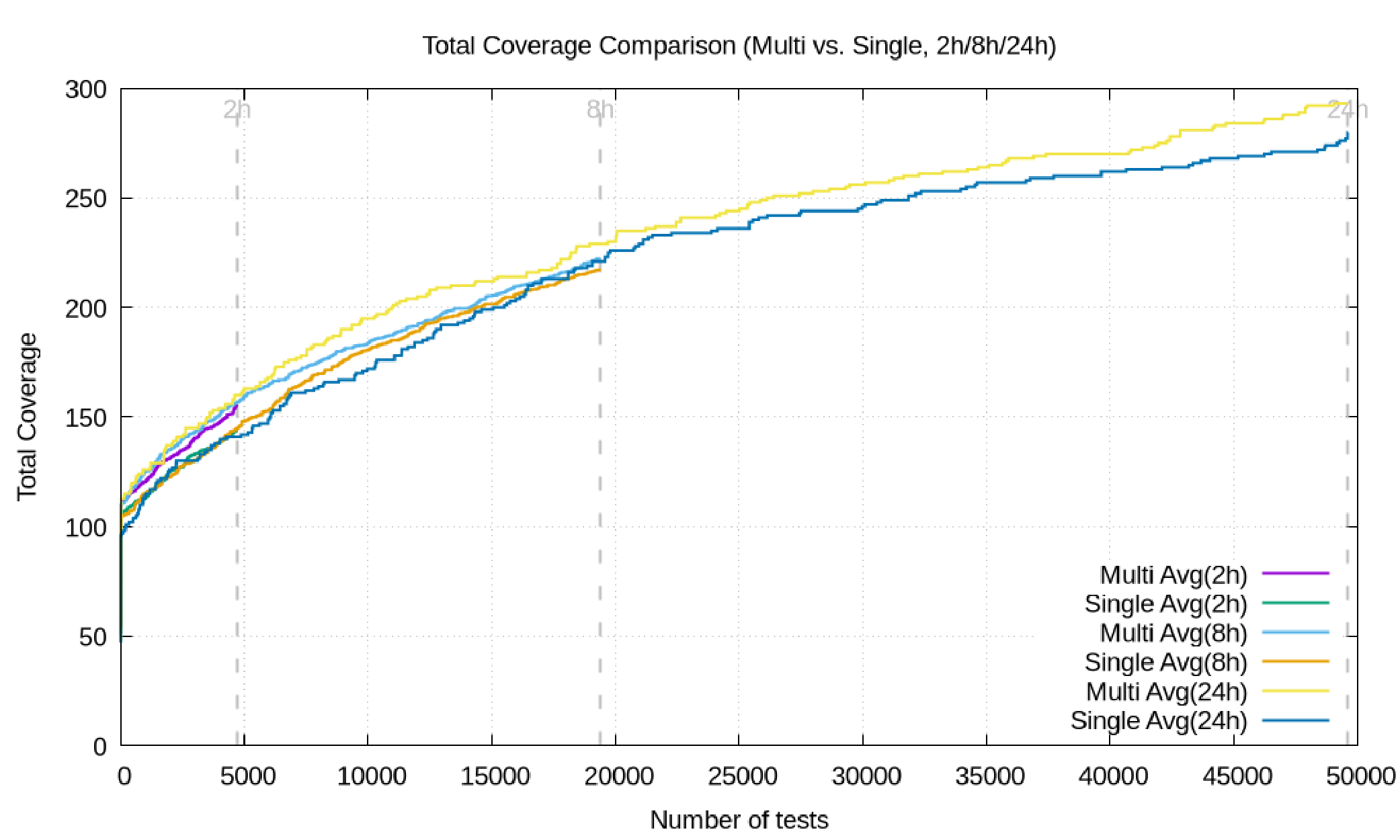}
\caption[
Total coverage Multi vs. Single 2h/8h/24h running results merged
]{
Total coverage of the Multi and Single configurations, 2h/8h/24h running results merged. Vertical dashed lines indicate the execution boundaries corresponding to 2h, 8h, and 24h runs.
}
\label{fig:coverage-multi-vs-single-2h-8h-24h}
\end{figure}

From \autoref{fig:coverage-multi-vs-single-2h} and \autoref{fig:coverage-multi-vs-single-8h}, the following trend can be observed. In short fuzzing runs, \texttt{multi} exhibits higher exploration efficiency and is able to discover more code coverage than \texttt{single}. 
However, when fuzzing is executed for a longer period of time, the difference in the number of detected coverage elements between \texttt{multi} and \texttt{single} gradually decreases.
This behavior can be interpreted as reflecting the difference in exploration efficiency between the proposed method (\texttt{multi}) and the conventional method (\texttt{single}), as discussed by Liyanage et al.~\cite{liyanage2023reachable}.
In addition, as a reference, we conducted a single 24-hour fuzzing campaign run, the results of which are shown in \autoref{fig:coverage-multi-vs-single-24h}. 
To ensure a fair comparison between \texttt{multi} and \texttt{single}, the analysis was limited to the first 49,612 test executions in both cases.
\autoref{fig:coverage-multi-vs-single-2h-8h-24h} combines the results of the 2h, 8h, and 24h experiments, showing that \texttt{multi} discovers new
coverage more efficiently than \texttt{single} in the early stages of fuzzing campaign.
The functions targeted for the measurement of code coverage in this experiment are listed in \autoref{list:linux-kernel-target-list}, \autoref{list:optee-os-target-list}, and \autoref{fig:coverage-multi-vs-single-24h} in Appendix \ref{Appendix: Additional Details} (\nameref{Appendix: Additional Details}). 
The functions contained in the files under these directories are treated as coverage measurement targets.

\FloatBarrier

\subsection{QEMU performance overhead results}
In the experiment, the number of CPUs allocated to the Docker container was set to one. In addition, a P-core was assigned on the \texttt{i7-14000} platform(\texttt{Env2}).

To summarize the results, this experiment aimed to quantify the performance overhead introduced by \textit{MTCFuzz} on the baseline QEMU. 
The results for \texttt{Env1} are shown in \autoref{tab:hackbench_result_9700k}, and those for \texttt{Env2} are presented in \autoref{tab:hackbench_result_14000}. 
In the process mode, both \texttt{Env1} and \texttt{Env2} exhibited a median overhead increase of approximately 45–65\%. 
In the thread mode, \texttt{Env1} showed an overhead increase of approximately 50\%, while \texttt{Env2} exhibited a substantially larger increase of approximately 83\%. 
\autoref{tab:qemu-performance-by-cpu} summarizes the results for each execution environment. 
This figure shows that the overhead varies depending on the CPU. Although the overhead is smaller on the \texttt{i7-9700K} than on the \texttt{i7-14000}, 
the runtime performance of the \texttt{i7-14000} is overall better. This can be attributed to the higher computational capability of the \texttt{i7-14000} compared with the \texttt{i7-9700K}, 
which allows it to perform more computations; however, this also leads to a larger overhead associated with recording coverage information. 
A larger overhead was observed in \texttt{Env2} compared to \texttt{Env1}. This can be attributed to the higher computational performance of the \texttt{i7-14000} used in
\texttt{Env2} relative to the \texttt{i7-9700K}. Consequently, the impact of bottleneck components becomes more pronounced on \texttt{i7-14000}.

\begin{figure}[th]
\centering
\includegraphics[width=\hsize]{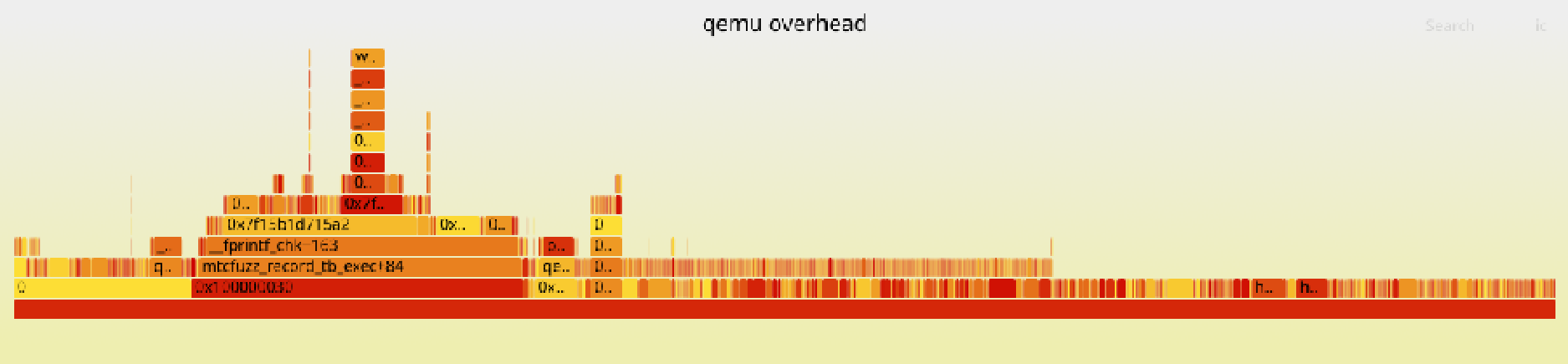}
\caption[Visualized qemu overhead]{Visualization of stack frames for QEMU execution overhead.
This figure shows a stack frame visualization aggregated from QEMU's stack traces collected during the execution of \texttt{hackbench}. Each frame represents an executed function, and the width of a frame indicates the relative proportion of execution time spent in that function.}
\label{fig:visualized-qemu-overhead}
\end{figure}

During hackbench execution, we measured the execution behavior of the QEMU using the \texttt{bpftrace} script shown in \autoref{code:bpftrace-profile} on the \texttt{Env2} 
host machine and aggregated the sampled stack traces into a stack flame graph. The result is shown in \autoref{fig:visualized-qemu-overhead}. 
From this stack flame graph, we observed that the call sequence from \texttt{mtcfuzz\_record\_tb\_exec} to \texttt{\_\_fprintf\_chk} appears four times among the top-ranked execution paths.
This indicates that the execution of \texttt{fprintf} for trace logging introduces a significant performance overhead.

\begin{lstlisting}[
    language=C, 
    float=th, 
    caption={bpftrace on-CPU profile script},
    label={code:bpftrace-profile},
    numbers=left,
    basicstyle=\ttfamily\normalsize,
    breaklines=true
]
profile:hz:99 /pid == $1/ {
  @[ustack] = count();
}

END {
  print(@);
}
\end{lstlisting}

\begin{table}[th]
  \centering
  \caption{hackbench: Process and Thread Performance (Env1)}
  \label{tab:hackbench_result_9700k}
  \resizebox{\textwidth}{!}{%
  \begin{tabular}{lcccccccc}
    \toprule
    & \multicolumn{4}{c}{Process} & \multicolumn{4}{c}{Thread} \\
    \cmidrule(lr){2-5} \cmidrule(lr){6-9}
    QEMU Type
      & Min & Max & Mean & Median
      & Min & Max & Mean & Median \\
    \midrule
    Baseline QEMU
      & 28.343 & 29.324 & 28.710 & 28.638
      & 23.062 & 27.886 & 23.633 & 23.292 \\
    MTCFuzz QEMU
      & 41.236 & 43.062 & 41.743 & 41.644
      & 34.677 & 39.488 & 35.546 & 35.084 \\
    \midrule
    Overhead (\%)
      & 45.5 & 46.8 & 45.4 & 45.5
      & 50.3 & 41.6 & 50.4 & 50.6 \\
    \bottomrule
  \end{tabular}}
\end{table}

\begin{table}[th]
  \centering
  \caption{hackbench: Process and Thread Performance (Env2)}
  \label{tab:hackbench_result_14000}
  \resizebox{\textwidth}{!}{%
  \begin{tabular}{lcccccccc}
    \toprule
    & \multicolumn{4}{c}{Process} & \multicolumn{4}{c}{Thread} \\
    \cmidrule(lr){2-5} \cmidrule(lr){6-9}
    QEMU Type
      & Min & Max & Mean & Median
      & Min & Max & Mean & Median \\
    \midrule
    Baseline QEMU
      & 14.229 & 14.586 & 14.381 & 14.365
      & 11.138 & 13.293 & 11.471 & 11.337 \\
    MTCFuzz QEMU
      & 21.242 & 26.978 & 23.668 & 23.761
      & 18.104 & 24.647 & 21.308 & 20.807 \\
    \midrule
    Overhead (\%)
      & 49.3 & 85.0 & 64.5 & 65.4
      & 62.5 & 85.4 & 85.8 & 83.5 \\
    \bottomrule
  \end{tabular}}
\end{table}

\begin{table}[th]
  \centering
  \caption{Performance and Overhead Comparison by CPU (hackbench process mode)}
  \label{tab:qemu-performance-by-cpu}
  \resizebox{\textwidth}{!}{%
  \begin{tabular}{lcccccccccccccccc}
    \toprule
    & \multicolumn{4}{c}{Process (s)}
    & \multicolumn{4}{c}{Thread (s)}
    & \multicolumn{4}{c}{Process OH (\%)}
    & \multicolumn{4}{c}{Thread OH (\%)} \\
    \cmidrule(lr){2-5}
    \cmidrule(lr){6-9}
    \cmidrule(lr){10-13}
    \cmidrule(lr){14-17}
    CPU &
    Min & Max & Mean & Median &
    Min & Max & Mean & Median &
    Min & Max & Mean & Median &
    Min & Max & Mean & Median \\
    \midrule
    i7-9700K &
      28.343 & 29.324 & 28.710 & 28.638 &
      23.062 & 27.886 & 23.633 & 23.292 &
      45.5 & 46.8 & 45.4 & 45.5 &
      50.3 & 41.6 & 50.4 & 50.6 \\
    i7-14000 &
      14.229 & 14.586 & 14.381 & 14.365 &
      11.138 & 13.293 & 11.471 & 11.337 &
      49.3 & 85.0 & 64.5 & 65.4 &
      62.5 & 85.4 & 85.8 & 83.5 \\
    \bottomrule
  \end{tabular}}
\end{table}

\FloatBarrier

\subsection{Address filter performance results}

The performance of the filtering process was evaluated by increasing the number of filters from $2^{1}$ to $2^{15}$ and repeating this test three times. The results are shown in \autoref{tab:filter_test_results}. The number of test executions ranged from 441 to 449 in Test~01, from 441 to 449 in Test~02, and from 444 to 451 in Test~03, regardless of the number of filters. As shown in \autoref{tab:filter_test_statistics}, the standard deviation in all tests remained around 2, indicating that this variation can be attributed to the noise of the measurement at runtime. These results confirm that the performance of the filtering process is not significantly affected even as the number of filters increases.

Since the filtering algorithm has a time complexity of $O(M \cdot \log_{2} N)$, a filtering scale that would cause a notable performance impact for example, doubling the execution time would require a filter count of the order of $2^{30} = 1{,}073{,}741{,}824$. However, this corresponds to an extremely large number of symbols for software typically targeted in fuzzing. In such a case, rather than configuring a large number of fine-grained filters, it would be simpler and more practical to apply coarse-grained filtering, such as using the entire address range of the Linux kernel and the entire address range of OpenSBI. This approach would, in turn, result in a smaller number of filters. Therefore, we conclude that the performance of the filtering mechanism in \textit{MTCFuzz} does not pose a practical limitation in realistic system environments.

\begin{table}[th]
    \centering
    \caption{Fuzzing Execution Counts under Different Filter Configurations}
    \label{tab:filter_test_results}
    \begin{tabular}{rccc}
        \toprule
        \textbf{Number of Filters} & \textbf{Test 01} & \textbf{Test 02} & \textbf{Test 03} \\
        \midrule
        2($2^{1}$)      & 441 & 441 & 451 \\
        4($2^{2}$)      & 443 & 445 & 448 \\
        8($2^{3}$)      & 445 & 445 & 448 \\
        16($2^{4}$)     & 448 & 448 & 449 \\
        32($2^{5}$)     & 449 & 443 & 448 \\
        64($2^{6}$)     & 443 & 447 & 446 \\
        128($2^{7}$)    & 442 & 444 & 449 \\
        256($2^{8}$)    & 443 & 447 & 449 \\
        512($2^{9}$)    & 441 & 446 & 447 \\
        1024($2^{10}$)  & 445 & 446 & 444 \\
        2048($2^{11}$)  & 446 & 441 & 449 \\
        4096($2^{12}$)  & 446 & 444 & 449 \\
        8192($2^{13}$)  & 448 & 445 & 445 \\
        16384($2^{14}$) & 444 & 446 & 446 \\
        32768($2^{15}$) & 441 & 449 & 445 \\
        \bottomrule
    \end{tabular}
\end{table}

\begin{table}[th]
    \centering
    \caption{Statistical Summary of Test Execution Counts}
    \label{tab:filter_test_statistics}
    \begin{tabular}{lcccc}
        \toprule
        \textbf{Test} & \textbf{Mean} & \textbf{Std.\ Dev.} & \textbf{Min} & \textbf{Max} \\
        \midrule
        Test01 & 444.33 & 2.57 & 441 & 449 \\
        Test02 & 445.13 & 2.22 & 441 & 449 \\
        Test03 & 447.53 & 1.89 & 444 & 451 \\
        \bottomrule
    \end{tabular}
\end{table}

\FloatBarrier

\subsection{Snapshot load performance results}

The results of the test executions with and without snapshot loading are shown in \autoref{tab:snapshot-load-performance}. Across all five runs, disabling snapshot loading results in an approximately $12\%$ improvement in execution throughput. In the current implementation of \textit{MTCFuzz}, a snapshot is loaded before each test execution in order to reset the system state and eliminate the effects of previously executed test cases. Considering this design choice, an execution throughput overhead of approximately $12\%$ introduced by snapshot loading is within an acceptable range.

\begin{table}[th]
\centering
\caption{Comparison of test case execution counts with and without snapshot loading}
\label{tab:snapshot-load-performance}
\resizebox{\textwidth}{!}{%
\begin{tabular}{lccc}
\toprule
 & Snapshot load enabled & Snapshot load disabled & Performance improvement rate (\%) \\
\midrule
Test01 & 449 & 507 & 12.92 \\
Test02 & 448 & 506 & 12.95 \\
Test03 & 448 & 505 & 12.72 \\
Test04 & 447 & 503 & 12.53 \\
Test05 & 448 & 503 & 12.28 \\
\midrule
Average & 448.0 & 504.8 & 12.68 \\
\bottomrule
\end{tabular}}
\end{table}

\FloatBarrier

\subsection{Fuzzing OP-TEE cryptographic API}
\label{sec:cve-finding-result}

As a result of this evaluation, we discovered a vulnerability in the  optee driver in the Linux kernel that is caused by insufficient validation of shared memory sizes.
Specifically, in the error-handling path of the routine that sets up shared memory based on the buffer and buffer size passed to the API.
Due to an issue in the cleanup logic for shared memory during error handling,
a NULL pointer dereference leads to kernel crash. The kernel panic log observed during this reproduction is shown in \autoref{code:kernel-oops-CVE-2025-40031}.

In our experiments, we used Linux kernel version 6.14.0-gb8a233c2155a.
We also verified that the vulnerability is reproducible in the Linux kernel version 6.17-rc5, 
which is the latest available mainline kernel version at the time of discovery.
We confirmed that the bug exists in the mainline kernel as well. 

Then, we reported the conditions under which this bug is triggered, along with reproduction steps and a reproduction script, to the OP-TEE development mailing list.
We also tested the bug fix patch developed by the maintainers and cooperated in the validation of the bug fix~\cite{trustedfirmwareBUGTee_shm}.
The vulnerability was quickly fixed, and after the fix was merged into the mainline Linux kernel, it was assigned to CVE-2025-40031~\cite{CVE-2025-40031}.

\begin{lstlisting}[
    language=C, 
    float=th, 
    caption={Kernel OOPS log by CVE-2025-40031.},
    label={code:kernel-oops-CVE-2025-40031},
    numbers=left,
    basicstyle=\ttfamily\scriptsize,
    breaklines=true
]
[   16.143987] Unable to handle kernel NULL pointer dereference at virtual address 0000000000000008
[   16.144141] Mem abort info:
[   16.144215]   ESR = 0x0000000096000004
[   16.144246]   EC = 0x25
** replaying previous printk message **
[   16.144246]   EC = 0x25: DABT (current EL), IL = 32 bits
[   16.144271]   SET = 0, FnV = 0
[   16.144289]   EA = 0, S1PTW = 0
[   16.144308]   FSC = 0x04: level 0 translation fault
[   16.144325] Data abort info:
[   16.144335]   ISV = 0, ISS = 0x00000004, ISS2 = 0x00000000
[   16.144346]   CM = 0, WnR = 0, TnD = 0, TagAccess = 0
[   16.144358]   GCS = 0, Overlay = 0, DirtyBit = 0, Xs = 0
[   16.144412] user pgtable: 4k pages, 52-bit VAs, pgdp=0000000048b34b00
[   16.144432] [0000000000000008] pgd=0800000040bb3403, p4d=0000000000000000
[   16.144876] Internal error: Oops: 0000000096000004 [#1]  SMP
[   16.146429] Modules linked in:
[   16.146775] CPU: 0 UID: 0 PID: 148 Comm: xtest Not tainted 6.17.0-rc5 #58 PREEMPT
[   16.146995] Hardware name: linux,dummy-virt (DT)
[   16.147181] pstate: 21402005 (nzCv daif +PAN -UAO -TCO +DIT -SSBS BTYPE=--)
[   16.147330] pc : unpin_user_pages+0x78/0xd0
[   16.147763] lr : unpin_user_pages+0xa0/0xd0
[   16.147842] sp : ffff800084403d20
[   16.147912] x29: ffff800084403d20 x28: fff00000054aa300 x27: 0000000000000000
[   16.148089] x26: 0000000000000000 x25: 0000000000000000 x24: 0000000000000000
[   16.148235] x23: fff00000004fb5a8 x22: 0000000000000001 x21: 000000000000000d
[   16.148401] x20: fff0000000b2f9c0 x19: 0000000000000011 x18: 0000000000000001
[   16.148544] x17: 0000000000000000 x16: 0000000000000000 x15: 0000000000000000
[   16.148659] x14: 0000000000000002 x13: 0000000000000002 x12: 0000000000037d0f
[   16.148786] x11: fff0000001dad708 x10: 000000000000003f x9 : 0000000000000d1b
[   16.148925] x8 : 00000000000007e0 x7 : 0000000000000001 x6 : 000000000000000d
[   16.149039] x5 : ffffffffffffffff x4 : ffffffffffffffff x3 : 000000000000000e
[   16.149167] x2 : 0000000000000000 x1 : 0000000000000000 x0 : ffffc1ffc0fd68c0
[   16.149351] Call trace:
[   16.149520]  unpin_user_pages+0x78/0xd0 (P)
[   16.149684]  tee_shm_put+0x134/0x184
[   16.149783]  tee_shm_fop_release+0x14/0x24
[   16.149866]  __fput+0xcc/0x2dc
[   16.149925]  fput_close_sync+0x40/0x108
[   16.149991]  __arm64_sys_close+0x38/0x7c
[   16.150058]  invoke_syscall+0x48/0x110
[   16.150127]  el0_svc_common.constprop.0+0x40/0xe8
[   16.150227]  do_el0_svc+0x20/0x2c
[   16.150303]  el0_svc+0x34/0xf0
[   16.150369]  el0t_64_sync_handler+0xa0/0xe4
[   16.150439]  el0t_64_sync+0x198/0x19c
[   16.150629] Code: aa0203e3 eb02027f 54000109 f8627a82 (f9400444)
[   16.150940] ---[ end trace 0000000000000000 ]---
[   16.151230] Kernel panic - not syncing: Oops: Fatal exception
[   16.151466] SMP: stopping secondary CPUs
[   16.151838] Kernel Offset: disabled
[   16.151911] CPU features: 0x000000,0000d180,2bbe33e1,957e7f3f
[   16.152019] Memory Limit: none
[   16.152284] ---[ end Kernel panic - not syncing: Oops: Fatal exception ]---
\end{lstlisting}

\FloatBarrier

\section{Summary of Evaluation}

In this chapter, we evaluated the effectiveness of the proposed method in terms of code
exploration capability, the QEMU overhead introduced by code coverage measurement,
the processing performance of the address filtering mechanism, and the overhead caused
by restoring snapshots. In addition, we successfully discovered a new CVE by performing fuzzing
against the OP-TEE API.

The evaluation results demonstrated that the proposed method achieves higher code
coverage than conventional CGF that targets only a single software component, and is able
to discover a larger number of execution paths within a limited time.
Furthermore, the performance overhead introduced by extending QEMU to support
code coverage measurement, as well as the cost associated with the address filtering
mechanism, remained within a practical range, confirming that the proposed method is
feasible as a practical fuzzing approach.

From these results, we conclude that the proposed method, which integrates and utilizes
the code coverage of multiple software components, is effective in improving exploration
efficiency in cooperative execution environments where conventional approaches face
difficulties.

\chapter{Discussion}
\label{ch:discussion}
Based on the evaluation results obtained in chapter \ref{ch:Evalation}, this chapter discusses the effectiveness and limitations of the proposed approach.

\section{Effectiveness of the Proposed Approach}

In sections \ref{lab:Coverage-Evaluation-of-OpenSBI-Base-Extension-result} and \ref{lab:CGF-with-Multi-Target-Coverage-Feedback-result}, we evaluated the code coverage exploration capability, and in both experiments, the proposed method achieved higher code coverage than the existing approach. This result can be interpreted as follows. 
In conventional CGF, when execution crosses software boundaries, for example, when execution transitions from the Linux kernel to OpenSBI due to the \texttt{ecall}, even if no new coverage is discovered on the Linux kernel side, new coverage may be reached on the OpenSBI side. 
However, because CGF targeting only the Linux kernel does not utilize the coverage information obtained in OpenSBI, \textsc{IsInteresting} is likely to judge that the executed test case did not yield useful results. 
In contrast, the proposed method continues the coverage measurement even when execution paths transition from the Linux kernel to OpenSBI. Therefore, coverage that is previously invisible to conventional CGF can now be utilized, and even in cases where no new coverage is discovered on the Linux kernel side, \textsc{IsInteresting} can judge the test case as beneficial if new coverage is detected in OpenSBI. 
For this reason, we consider that the ability of the proposed method to measure coverage across software boundaries contributed to its higher coverage exploration capability compared with conventional CGF.

The proposed method enables fuzzing in architectures where multiple software components operate cooperatively. Therefore, it is particularly effective in software systems whose functionality is not confined to a single software component, such as environments in which the kernel and the firmware operate in cooperation.

\section{Limitations}

\textit{MTCFuzz} leverages QEMU as the fuzzing execution environment,
which enables comprehensive monitoring of the behavior of all software
running inside QEMU.
However, the current design assumes that only a single fuzzing target
environment is executed within a QEMU instance.
Although it is technically possible to emulate multiple systems by running
nested QEMU instances inside the Linux environment launched by \textit{MTCFuzz},
the current implementation does not support such configurations.

\section{Challenges}

There are two main challenges in \textit{MTCFuzz}. 
The first challenge, as demonstrated in the experiments in the Chapter~\ref{ch:Evalation},
is that the overhead of code coverage measurement using QEMU is relatively high. 
The second challenge is that the creation of test harnesses must be carried out by
the fuzzing practitioner.

\vskip.5\baselineskip
\noindent\textbf{Improving the performance of QEMU-based coverage measurement:}

Execution efficiency is essential for fuzzing and reducing overhead is an important issue.
In the current implementation, \texttt{fprintf} is invoked every time 
\texttt{mtcfuzz\_record\_tb\_exec} is called, and the \texttt{bpftrace} profiling results
showed that this routine constitutes a major bottleneck.
To mitigate this bottleneck, the following improvements can be considered:

\begin{enumerate}
\item Store coverage data in memory until writing to a file in order to reduce
      the number of \texttt{fprintf} executions.
\item Replace file I/O with shared memory communication between the fuzzer
      and QEMU to eliminate the overhead of \texttt{fprintf} entirely.
\end{enumerate}

In the case of approach (1), the question is how frequently the buffered data 
should be flushed. As shown in Section~\ref{sec:OpenSBI-Base-Extension-coverage},
one execution of the test harness records approximately $35{,}000$ addresses on average.
Since the output is written as text rather than binary, each 64-bit value must be
converted to a hexadecimal string (up to 16 characters), and with the \texttt{0x}
prefix and newline, one entry requires up to 19 bytes. If a 4096-byte buffer is used,
writing the data whenever approximately 200 entries accumulate would result in only
175 write operations per run, reducing the number of \texttt{fprintf} calls by
approximately $99.5\%$. Therefore, this approach has the potential to significantly
reduce the measurement overhead.
In the case of approach (2), file I/O can be eliminated completely, removing the
\texttt{fprintf} overhead entirely. However, when using shared memory, QEMU must
track how much data has been written, while the fuzzer must manage how much has been
consumed. If QEMU overwrites the buffer before the fuzzer finishes reading it,
the recorded coverage information becomes unreliable. Therefore, although this
approach removes the \texttt{fprintf} overhead, it requires careful buffer management
between QEMU and the fuzzer.

\vskip.5\baselineskip
\noindent\textbf{Improving test harness creation:}

Although \textit{MTCFuzz} provides a fuzzing framework, the creation of test harnesses
must be performed by the user. Since the primary fuzzing targets of \textit{MTCFuzz}
include kernels, firmware, Trusted OSs, and Trusted Applications, test harnesses must
be implemented using the APIs provided by these components.
Unlike file-format fuzzing (e.g., PNG or ELF), where valid initial seeds can be
used and mutated, API fuzzing requires knowledge of execution dependencies, 
parameter preparation, and ordering of API invocations.
For example, the \texttt{write} system call requires a file descriptor (fd) as its
first argument, which must be obtained beforehand using \texttt{open}. Furthermore,
the fd may refer to a regular file, a pseudo file, or even a network socket.
In addition, the execution state can vary depending on operations such as modifying
the file offset via \texttt{lseek} or holding a lock using \texttt{flock} in another
thread before \texttt{write} is executed. Consequently, constructing appropriate
test harnesses for API-based fuzzing is not a trivial task.
Therefore, although \textit{MTCFuzz} provides a framework for Multi-target fuzzing,
the burden of test harness development remains a significant challenge. Addressing
this issue would further enhance the practicality and efficiency of fuzzing using
the proposed approach.

\section{Summary of Discussion}
In this chapter, we discussed the effectiveness and limitations of the proposed approach.
Based on the evaluation results in Chapter~\ref{ch:Evalation}, the proposed method was
shown to enable code coverage measurement across software boundaries in systems where
multiple software components operate cooperatively. As a result, it expands the range
of code exploration compared to conventional CGF and enables more effective exploration.
At the same time, we identified several practical challenges, including the overhead
of QEMU-based coverage measurement and the burden of creating test harnesses.

Overall, the proposed approach is both meaningful and practical, although there remains
room for improvement. These findings form the basis for the future enhancements and
extensions that will be addressed in Chapter~\ref{ch:future-work}.

\chapter{Related Work}
\label{ch:related-work}
In this chapter, we discuss related work based on two key aspects of the proposed approach:
its implementation as a QEMU-based fuzzing framework and its focus on system software such
as kernels and Trusted OS.

\vskip.5\baselineskip
\noindent\textbf{QEMU-based Fuzzing Frameworks:}
As studies that employ QEMU as the execution environment for fuzzing, Schumilo et al.
proposed kAFL~\cite{schumilo2017kafl}, and Malmain et al. proposed
LibAFL QEMU~\cite{malmain2024libafl}.
The kAFL measures code coverage using hardware support, namely Intel Processor
Trace (Intel PT). Since coverage measurement is performed using hardware tracing,
the overhead is extremely low, reported to be less than $5\%$~\cite{schumilo2017kafl}.
It supports fuzzing of operating systems such as Linux and Windows. Because coverage
is obtained via Intel PT, software modification (e.g., rebuilding with instrumentation)
is not required. Basic blocks that should be excluded from coverage measurement can be
registered in a blacklist so that they are ignored. As kAFL relies on Intel PT,
the host machine must be equipped with an Intel Skylake or later processor. In addition,
the host operating system must be Ubuntu 20.04 or later, or Debian Bullseye or later,
with a Linux kernel customized for kAFL. Although QEMU is used as the fuzzing
execution environment, the supported target architectures are limited to x86 and x86-64,
and architectures such as ARM and RISC-V are not supported. Furthermore,
kAFL does not adopt snapshot restoration because it is considered costly; instead,
it continuously uses the same VM instance, and non-deterministic behavior such as interrupts
is filtered using the blacklist mechanism.

The LibAFL QEMU is a fuzzing framework implemented as part of LibAFL~\cite{fioraldi2022libafl}. 
The LibAFL QEMU customizes QEMU so that QEMU, which normally operates as a standalone application, can be embedded and controlled as
a library from the fuzzer. The code coverage measurement is performed inside QEMU, and coverage
is collected at the basic block granularity in TCG. While kAFL allows users to
configure coverage exclusion using a blacklist, LibAFL QEMU enables specifying
coverage regions using an allow-list mechanism. It also incorporates several techniques
to support efficient snapshot restoration: 1) device state management relies on existing mechanisms in QEMU, 2) memory snapshots are restored using differential state management, 3) block
devices are launched in read-only mode while write operations are redirected to a shadow
region in RAM. As a result, reverting a snapshot can be completed simply by discarding the
shadow region. The operating systems and CPU architectures that can be fuzzed with
LibAFL QEMU follow those supported by QEMU, and no software rebuilding is required
for coverage measurement.

In both kAFL and LibAFL QEMU, test suites must be implemented by the
developer who performs fuzzing.

\vskip.5\baselineskip
\noindent\textbf{Fuzzing for Kernel/Trusted OS:}
syzkaller is a fuzzer that targets multiple operating systems, including
Linux, FreeBSD, and Windows. It supports QEMU as an execution environment;
however, syzkaller uses upstream or distribution provided QEMU as-is
and does not require any customization of QEMU. In addition to QEMU-based
execution, fuzzing can also be performed on physical hardware. When fuzzing the
Linux kernel, code coverage is measured using KCOV~\cite{kernelKCOVCode},
and therefore the target Linux kernel must be built with \texttt{KCOV} enabled.
One of the notable features of syzkaller is its domain-specific language, syzlang. 
Using syzlang, system call specifications are defined,
and based on these specifications, syzkaller automatically generates valid
system call invocation sequences and corresponding test programs. Although developers
are required to add new system call specifications manually, a large number of
definitions are already available in the official repository, allowing practical
fuzzing to be performed immediately after cloning and building the source code.
In addition to system calls provided by the kernel, pseudo-syscalls can also be
defined~\cite{pwningTicklingKsmbd}. Using these specifications as input,
syzkaller constructs test programs and mutates them during execution,
efficiently expanding code coverage.

Although syzkaller is already a highly capable fuzzer, many approaches
have been proposed based on it. For example, SyzDirect~\cite{tan2023syzdirect}
performs \textit{Directed Greybox Fuzzing} for the Linux kernel, and
SyzParam~\cite{sun2025syzparam} focuses on fuzzing device drivers via
sysfs modifiable parameters.
Furthermore, SyzTrust, proposed by Wang et al., is a fuzzer targeting
Trusted Execution Environments (TEE) and is also based on syzkaller.
SyzTrust focuses on fuzzing Trusted OSs and executes on real hardware
rather than QEMU. Since mechanisms such as KCOV are not available as in
Linux, coverage measurement is carried out using a debug probe.
To fuzz Trusted OS components, SyzTrust utilizes the GP TEE Internal
Core API, the specifications of which are described using syzlang.
API execution against the Trusted OS is triggered from a Trusted Application.

\begin{table}[th]
\centering
\caption{Comparison of system fuzzing frameworks}
\label{tab:related-work-comparison}
\resizebox{\textwidth}{!}{%
\begin{tabular}{lccc}
\hline
 & \textbf{kAFL} & \textbf{LibAFL QEMU} & \textbf{MTCFuzz} \\
\hline
Target CPU architecture
 & x86-64 / x86
 & QEMU-supported CPUs
 & QEMU-supported CPUs \\
Host OS
 & Ubuntu / Debian
 & Any
 & Any \\
Language
 & C
 & Rust
 & Python \\
Performance
 & $\checkmark$
 & $(\checkmark)$
 & $(\checkmark)$ \\
Instrumentation Method
 & Hardware-assisted (Intel PT)
 & Emulator-based (QEMU)
 & Emulator-based (QEMU) \\
Target Model
 & Single-target
 & Single-target
 & Multi-target \\
Snapshot Save / Restore
 & No
 & Yes
 & Yes \\
State Reset Mechanism
 & Reset
 & Snapshot
 & Snapshot \\
Representative Use Case
 & Kernel fuzzing
 & Binary / kernel fuzzing
 & OS--Firmware cooperative fuzzing \\
\hline
\end{tabular}}
\end{table}

\vskip.5\baselineskip
\noindent\textbf{Summary and Positioning:}
\autoref{tab:related-work-comparison} summarizes the key differences between kAFL,
LibAFL, and our proposed system, \textit{MTCFuzz}.
In summary, existing QEMU-based fuzzing frameworks focus on maximizing the
capabilities of QEMU and introducing necessary customizations in order to perform
efficient fuzzing. Meanwhile, kernel-oriented fuzzers support efficient test
generation by enabling developers to construct test harnesses using appropriate
combinations of system calls and APIs. However, these frameworks and fuzzers are
fundamentally designed for Single-target fuzzing: while they are highly effective
within their respective software domains, they do not provide unified Multi-target
coverage feedback across cooperatively executing software components.
Our proposed method addresses this limitation by enabling integrated coverage
measurement across software boundaries.

\FloatBarrier

\chapter{Future Work}
\label{ch:future-work}
As future work, we plan to integrate the proposed method into LibAFL QEMU, one of the state-of-the-art fuzzing frameworks. While the proof-of-concept implementation \textit{MTCFuzz} provides only the minimum necessary fuzzing functionality, LibAFL QEMU offers advanced capabilities such as high-performance snapshot mechanisms, flexible execution control, and support for modern fuzzing techniques including \textit{cmplog}~\cite{fioraldi2020weizz}. 
By realizing our approach—measuring and leveraging code coverage across multiple software boundaries—on top of LibAFL QEMU, we expect to extend conventional Single-target fuzzing workflows to cooperative execution environments such as kernel–firmware and kernel–Trusted OS systems. 
Furthermore, this integration is expected to facilitate broader adoption of our approach and enable large-scale, realistic evaluations using the latest fuzzing ecosystem.

\chapter{Conclusion}
\label{ch:conclusion}

This thesis addressed the challenge of fuzzing environments in which multiple
software components operate cooperatively, such as operating systems with
firmware or Trusted OS. Conventional fuzzing techniques primarily focus on a
single software component and have not sufficiently explored the unified use of
code coverage information in such cooperative execution environments.

To overcome this limitation, we proposed a method \textit{Multi-target Coverage-based Greybox Fuzzing} that leverages QEMU to enable
unified execution measurement of multiple software components running within a
virtualized environment. To demonstrate the practicality of the proposed
approach, we developed a prototype implementation, \textit{MTCFuzz}, and
evaluated it in environments where multiple software components operate in
coordination.

The evaluation results showed that the proposed method achieved higher code
coverage than conventional CGF.
Furthermore, the runtime overhead introduced by coverage measurement remained
within a practical range. These results indicate that the proposed method can
effectively enhance code exploration in environments where multiple software
components execute cooperatively.

This study demonstrates the feasibility of fuzzing multiple software components
in an integrated manner and provides a foundation for future research aimed at
improving performance, enhancing automation, and expanding applicability to a
broader range of systems.

\begin{acknowledgment}
I would like to express my deepest gratitude to Professors Ariyasu Suzaki,
Eiji Kuwana, and Atsuhiro Goto of the Institute of Information Security
for their invaluable guidance and support throughout this research.
I am also deeply grateful to the members of the Suzaki Laboratory and
the Institute of Information Security for their helpful advice,
information sharing, and continuous encouragement.
I would further like to thank my colleagues at my company for their
understanding, support that contributed to this research.

\end{acknowledgment}

\addtocontents{toc}{\protect\contentsline{chapter}{\bibname}{47}{}}

\begingroup
\renewcommand{\addcontentsline}[3]{}
\begin{bib}[100]
\bibliography{main}

\end{bib}
\endgroup





\appendix
\chapter{Terminology}

\begin{itemize}
  \item \textbf{Fuzzing}:
  A software testing technique that executes a program with numerous automatically generated inputs to uncover crashes or vulnerabilities.

  \item \textbf{Seed}:
  An input used as a template for generating new test cases through mutation during fuzzing.

  \item \textbf{Initial seed}:
  A seed input explicitly provided by the user at the start of a fuzzing campaign, which serves as the initial template from which subsequent seeds are generated.

  \item \textbf{Fuzzing campaign}:
  A continuous fuzzing execution session performed under fixed testing conditions.

  \item \textbf{OpenSBI}:
  RISC-V firmware that provides Supervisor Binary Interface (SBI) services to the operating system.

  \item \textbf{OP-TEE}:
  Trusted execution environment software running in the Secure World of ARM TrustZone.

  \item \textbf{TrustZone}:
  ARM security architecture that separates execution environments into Secure World and Normal World.

  \item \textbf{Basic block}:
  A straight-line sequence of instructions without internal branching.

  \item \textbf{Code coverage}:
  A measure of which parts of a program have been executed during testing.
  
  \item \textbf{Test case}:
An input executed on the software under test (SUT) during fuzzing.
In coverage-guided fuzzing, promising test cases are retained and used as new seeds.

\end{itemize}

\chapter{Additional Details}
\label{Appendix: Additional Details}

\begin{lstlisting}[
    language=c, 
    float=th, 
    caption={Linux kernel target list used in CGF with Multi-Target Coverage Feedback},
    label={list:linux-kernel-target-list},
    numbers=left,
    basicstyle=\ttfamily\footnotesize,
    breaklines=true
]
linux/drivers/tee
linux/drivers/tee/optee
\end{lstlisting}

\begin{lstlisting}[
    language=c, 
    caption={optee\_os target list used in CGF with Multi-Target Coverage Feedback},
    label={list:optee-os-target-list},
    numbers=left,
    basicstyle=\ttfamily\footnotesize,
    breaklines=true
]
optee_os/out/arm/core/tee
optee_os/out/arm/core/lib/zlib
optee_os/out/arm/core/lib/libtomcrypt
optee_os/out/arm/core/lib/libtomcrypt/src/mac/hmac
optee_os/out/arm/core/lib/libtomcrypt/src/mac/omac
optee_os/out/arm/core/lib/libtomcrypt/src/modes/xts
optee_os/out/arm/core/lib/libtomcrypt/src/modes/ctr
optee_os/out/arm/core/lib/libtomcrypt/src/modes/cbc
optee_os/out/arm/core/lib/libtomcrypt/src/modes/ecb
optee_os/out/arm/core/lib/libtomcrypt/src/math/fp
optee_os/out/arm/core/lib/libtomcrypt/src/math
optee_os/out/arm/core/lib/libtomcrypt/src/encauth/ccm
optee_os/out/arm/core/lib/libtomcrypt/src/ciphers/aes
optee_os/out/arm/core/lib/libtomcrypt/src/ciphers
optee_os/out/arm/core/lib/libtomcrypt/src/pk/ed25519
optee_os/out/arm/core/lib/libtomcrypt/src/pk/ecc
optee_os/out/arm/core/lib/libtomcrypt/src/pk/pkcs1
optee_os/out/arm/core/lib/libtomcrypt/src/pk/dsa
optee_os/out/arm/core/lib/libtomcrypt/src/pk/x25519
optee_os/out/arm/core/lib/libtomcrypt/src/pk/dh
optee_os/out/arm/core/lib/libtomcrypt/src/pk/rsa
optee_os/out/arm/core/lib/libtomcrypt/src/pk/asn1/oid
optee_os/out/arm/core/lib/libtomcrypt/src/pk/asn1/der/octet
optee_os/out/arm/core/lib/libtomcrypt/src/pk/asn1/der/sequence
optee_os/out/arm/core/lib/libtomcrypt/src/pk/asn1/der/set
optee_os/out/arm/core/lib/libtomcrypt/src/pk/asn1/der/generalizedtime
optee_os/out/arm/core/lib/libtomcrypt/src/pk/asn1/der/object_identifier
optee_os/out/arm/core/lib/libtomcrypt/src/pk/asn1/der/utf8
optee_os/out/arm/core/lib/libtomcrypt/src/pk/asn1/der/integer
optee_os/out/arm/core/lib/libtomcrypt/src/pk/asn1/der/ia5
optee_os/out/arm/core/lib/libtomcrypt/src/pk/asn1/der/bit
optee_os/out/arm/core/lib/libtomcrypt/src/pk/asn1/der/short_integer
optee_os/out/arm/core/lib/libtomcrypt/src/pk/asn1/der/printable_string
optee_os/out/arm/core/lib/libtomcrypt/src/pk/asn1/der/teletex_string
optee_os/out/arm/core/lib/libtomcrypt/src/pk/asn1/der/boolean
optee_os/out/arm/core/lib/libtomcrypt/src/pk/asn1/der/custom_type
optee_os/out/arm/core/lib/libtomcrypt/src/pk/asn1/der/utctime
optee_os/out/arm/core/lib/libtomcrypt/src/pk/asn1/der/choice
optee_os/out/arm/core/lib/libtomcrypt/src/pk/asn1/der/general
optee_os/out/arm/core/lib/libtomcrypt/src/pk/ec25519
optee_os/out/arm/core/lib/libtomcrypt/src/hashes
optee_os/out/arm/core/lib/libtomcrypt/src/hashes/sha2
optee_os/out/arm/core/lib/libtomcrypt/src/hashes/helper
optee_os/out/arm/core/lib/libtomcrypt/src/misc
optee_os/out/arm/core/lib/libtomcrypt/src/misc/ssh
optee_os/out/arm/core/lib/libtomcrypt/src/misc/pkcs5
optee_os/out/arm/core/lib/libtomcrypt/src/misc/crypt
optee_os/out/arm/core/lib/libtomcrypt/src/misc/pkcs12
optee_os/out/arm/core/lib/libtomcrypt/src/misc/base64
optee_os/out/arm/core/lib/libfdt
optee_os/out/arm/core/drivers/clk
optee_os/out/arm/core/drivers/rstctrl
optee_os/out/arm/core/drivers
optee_os/out/arm/core/drivers/gpio
optee_os/out/arm/core-lib/libmbedtls/mbedtls/library
optee_os/out/arm/core-lib/libutils/isoc
optee_os/out/arm/core-lib/libutils/isoc/arch/arm
optee_os/out/arm/core-lib/libutils/isoc/newlib
optee_os/out/arm/core-lib/libutils/compiler-rt/lib/builtins
optee_os/out/arm/core-lib/libutils/ext
optee_os/out/arm/core-lib/libutils/ext/arch/arm
optee_os/out/arm/lib/libunw
optee_os/out/arm/ta/avb
optee_os/out/arm/ta/trusted_keys
optee_os/out/arm/ta/remoteproc
optee_os/out/arm/ta/remoteproc/src
optee_os/out/arm/ta/pkcs11
optee_os/out/arm/ta/pkcs11/src
optee_os/out/arm/ta_arm64-lib/libmbedtls/mbedtls/library
optee_os/out/arm/ta_arm64-lib/libutils/isoc
optee_os/out/arm/ta_arm64-lib/libutils/isoc/arch/arm
optee_os/out/arm/ta_arm64-lib/libutils/isoc/newlib
optee_os/out/arm/ta_arm64-lib/libutils/compiler-rt/lib/builtins
optee_os/out/arm/ta_arm64-lib/libutils/ext
optee_os/out/arm/ta_arm64-lib/libutils/ext/arch/arm
optee_os/out/arm/ta_arm64-lib/libutee
optee_os/out/arm/ta_arm64-lib/libutee/arch/arm
optee_os/out/arm/ta_arm64-lib/libdl
\end{lstlisting}

\end{document}